\documentclass[fleqn,usenatbib]{mnras}

\usepackage{newtxtext,newtxmath}

\usepackage[T1]{fontenc}
\usepackage{ae,aecompl}

\usepackage{graphicx}	
\usepackage{physics}
\usepackage{subfigure}
\usepackage{xcolor}
\usepackage{txfonts}
\usepackage[normalem]{ulem}
\definecolor{MM}{HTML}{BF16ED}

\title[Eigenspectra: 3D Eclipse Mapping]{Eigenspectra: A Framework for Identifying Spectra from 3D Eclipse Mapping}

\author[M. Mansfield et al.]{
Megan Mansfield,$^{1}$\thanks{E-mail: meganmansifeld@uchicago.edu}
Everett Schlawin,$^{2}$
Jacob Lustig-Yaeger,$^{3,4}$
Arthur D. Adams,$^{5,6}$
\newauthor
Emily Rauscher,$^{6}$
Jacob Arcangeli,$^{7}$
Y. Katherina Feng,$^{8}$
Prashansa Gupta,$^{9}$
\newauthor
Dylan Keating,$^{10}$
Kevin B. Stevenson,$^{11}$
and Thomas G. Beatty$^{2}$
\\
$^{1}$Department of Geophysical Sciences, University of Chicago, 5734 S. Ellis Avenue, Chicago, IL 60637, USA\\
$^{2}$Department of Astronomy and Steward Observatory, University of Arizona, Tucson, AZ 85719, USA\\
$^{3}$Department of Astronomy and Astrobiology Program, University of Washington, Box 351580, Seattle, Washington 98195, USA\\
$^{4}$NASA Nexus for Exoplanet System Science, Virtual Planetary Laboratory Team, Box 351580, University of Washington, Seattle, Washington 98195, USA\\
$^{5}$Department of Astronomy, Yale University, New Haven, CT 06520, USA\\
$^{6}$Department of Astronomy, University of Michigan, Ann Arbor, MI 48109, USA\\
$^{7}$Anton Pannekoek Institute for Astronomy, University of Amsterdam, 1098 XH Amsterdam, The Netherlands\\
$^{8}$Department of Astronomy \& Astrophysics, University of California, Santa Cruz, CA 95064, USA\\
$^{9}$University of Montreal, Montreal, QC H3T 1J4, Canada\\
$^{10}$Department of Physics, McGill University, Montr\'eal, QC H3A 2T8, Canada\\
$^{11}$Johns Hopkins University Applied Physics Laboratory, Laurel, MD 20723, USA
}

\date{Accepted 2020 October 1. Received 2020 August 31; in original form 2020 May 20}

\pubyear{2020}

\begin{document}
\label{firstpage}
\pagerange{\pageref{firstpage}--\pageref{lastpage}}
\maketitle

\begin{abstract}
Planetary atmospheres are inherently 3D objects that can have strong gradients in latitude, longitude, and altitude. Secondary eclipse mapping is a powerful way to map the 3D distribution of the atmosphere, but the data can have large correlations and errors in the presence of photon and instrument noise. 
We develop a technique to mitigate the large uncertainties of eclipse maps by identifying a small number of dominant spectra to make them more tractable for individual analysis via atmospheric retrieval. We use the eigencurves method to infer a multi-wavelength map of a planet from spectroscopic secondary eclipse light curves. 
We then apply a clustering algorithm to the planet map to identify several regions with similar emergent spectra. We combine the similar spectra together to construct an ``eigenspectrum'' for each distinct region on the planetary map. 
We demonstrate how this approach could be used to isolate hot from cold regions and/or regions with different chemical compositions in observations of hot Jupiters with the \textit{James Webb Space Telescope} (\textit{JWST}). We find that our method struggles to identify sharp edges in maps with sudden discontinuities, but generally can be used as a first step before a more physically motivated modeling approach to determine the primary features observed on the planet.
\end{abstract}




\section{Introduction}

Planets are intrinsically 3D objects, but 1D models are often used to approximate planetary atmospheres. While 1D (vertical-only) models provide computationally inexpensive estimates of the vertical structure and radiative transfer in a single atmospheric column, temperature-pressure profiles can vary substantially for different locations around the planet. Inferring properties of 3D atmospheres with 1D models can correspondingly give biased abundance estimates \citep{feng2016nonUniform,Caldas2019}, since these models only approximate the arithmetic mean profile of what can be obtained by General Circulation Models \citep[GCMs;][]{blecic2017structure3Dw1Dretrieval}. 
On the other hand, GCMs have the ability to model atmospheric structures and dynamics fully three-dimensionally, but are much more demanding computationally, making them infeasible for inference and only able to best constrain the physics where robust data sets are available. 

Despite the necessity of 3D approaches to accurately model atmospheres, our current data from exoplanet atmospheres are almost entirely limited to 1D or 2D observations. 
For example, while phase curves are valuable in probing planetary brightness, they can reveal only longitudinal structure as the planet orbits its host star \citep[e.g.][]{knutson2007map189}, with vertical information also accessible if a phase curve is spectroscopic \citep{Stevenson2014}. In order to recover information about all three spatial dimensions, we must combine different data sets in ways that further exploit either the geometry of the system or the spectral imprints of the atmospheric structure.

Secondary eclipses of transiting planets offer valuable opportunities to observe and understand the multidimensional nature of exoplanets \citep[e.g.][]{Williams2006,rauscher2009eclipseMapping}.
As a planet goes behind its star, the stellar limb scans across the dayside hemisphere of the planet, permitting a 2D reconstruction of the planetary photosphere by probing the latitudinal structure. Combining these data with spectral information can add the third dimension, since in principle different wavelengths may probe different altitudes in the planet's atmosphere \citep[albeit not necessarily through a simple correspondence,][]{Dobbs2017}.

The first (and only) published eclipse map of a planet was for the hot Jupiter HD~189733b with the \textit{Spitzer Space Telescope} \citep{deWit2012eclipsemap189,majeau2012eclipsemap189}.
This map revealed a localized hot spot which was shifted eastward from the sub-stellar point, as predicted by GCMs \citep{showman2002circulation51peg} and found from the phase curve \citep{knutson2007map189}. The \textit{Spitzer} eclipse map, however, only probed the 2D structure at a single photospheric level because it used broadband photometry. It will be possible to construct spectroscopic secondary eclipse maps with the \textit{James Webb Space Telescope} (\textit{JWST}), which will allow investigations of changing atmospheric properties with altitude as well as with latitude and longitude. \textit{JWST} will provide unprecedented measurements of exoplanet atmospheres due to its large aperture and wavelength range for time series ($\sim 0.6$\,$\mu$m to $\sim 11$\,$\mu$m) \citep[e.g.][]{beichman2014pasp,greene2016jwst_trans,batalha2017pandexo}.
This could enable high precision eclipse mapping of virtually every bright hot Jupiter observed by \textit{JWST}.
Under the \citet{zhang2017bulkCompDynamics} analytic parameterization of atmospheric dynamics, for example, a single eclipse light curve of HD 189733b with NIRCam's F322W2 grism mode will localize the hot spot longitude to $\sim \pm 3.5 \deg$ in longitude \citep{schlawin2018JWSTforecasts}.

One of the challenges in mapping exoplanets is determining how to combine these pieces of spatial information in a way that extracts the maximum possible information on the physical state of the planet. 
Eclipse mapping does not provide a perfect proxy for each spatial dimension; one must account carefully for the inherent degeneracies and uncertainties when reconstructing a global brightness map. One approach is to assume a functional form of the variations in either temperature or flux in longitude and latitude, one that for example captures the angular dependence of the instellation and associated thermal energy budget \citep{Irwin2019}. From there one can calculate the associated molecular abundances and pressure-temperature profiles.

Another approach is to forgo any explicit parametrizations about the flux, and instead quantify the available information content from observations by constructing an orthogonal basis of light curves. Spherical harmonics represent an orthogonal basis for 2D representations of maps on a planet photosphere. However, the observations used to make an eclipse map are brightness as a function of time, and spherical harmonics are not orthogonal in this parameter space. \citet{Rauscher2018} addressed this issue by developing an orthogonal basis of eclipse light curves, 
which they term ``eigencurves,'' to best represent the information available from both phase variations and secondary eclipses at a single wavelength. 
They constructed these eigencurves from linear combinations of spherical harmonic maps. This approach avoids making a priori assumptions about the structure of brightness variations across the planet's photosphere while also providing 
the ability to directly assess the effects of orbital uncertainties on the retrieval. 

In order to extend this framework into the third dimension of multi-wavelength observations we present a method using K-means clustering to identify ``eigenspectra,''\footnote{ We acknowledge that the spectra are not orthogonal and therefore not formally ``eigenvectors,'' however we use this term colloquially in reference to the ``eigencurves'' used in our algorithm.} which are a set of spectra that together represent most of the variance in spectral properties observed over the dayside of the planet. The identification of these eigenspectra allows atmospheric retrievals to be performed on spectra with the smallest possible error bars, while ensuring that regions of the dayside with vastly different atmospheric properties do not get grouped into a single retrieval. We describe our method of identifying eigenspectra in Section~\ref{sec:methods}, using three hypothetical \textit{JWST} eclipse maps. We present the results of these hypothetical observations in Section~\ref{sec:results} and discuss the limitations of the eigenspectra method. 
We summarize our method and discuss its utility in mapping real planets in Section~\ref{sec:conclude}.

\section{Methodology}
\label{sec:methods}

To investigate the potential to perform 3D eclipse mapping with JWST, we first generate synthetic observations of an exoplanet observed in secondary eclipse at multiple wavelengths, and then attempt to recover our original planet map components and their respective emergent spectra. We describe our approach for generating synthetic eclipse light curves in Section~\ref{sec:construct}, followed by a description of our newly developed model for fitting these multi-dimensional data in Section~\ref{sec:extract}. The Python code developed for this paper is all publicly available on GitHub\footnote{\href{https://github.com/multidworlds/eigenspectra}{\texttt{eigenspectra}} code is available on GitHub.}. 

\subsection{Construction of Planet Maps and Eclipse Light Curves}
\label{sec:construct}

\begin{figure}
    \centering
    \includegraphics[width=\linewidth]{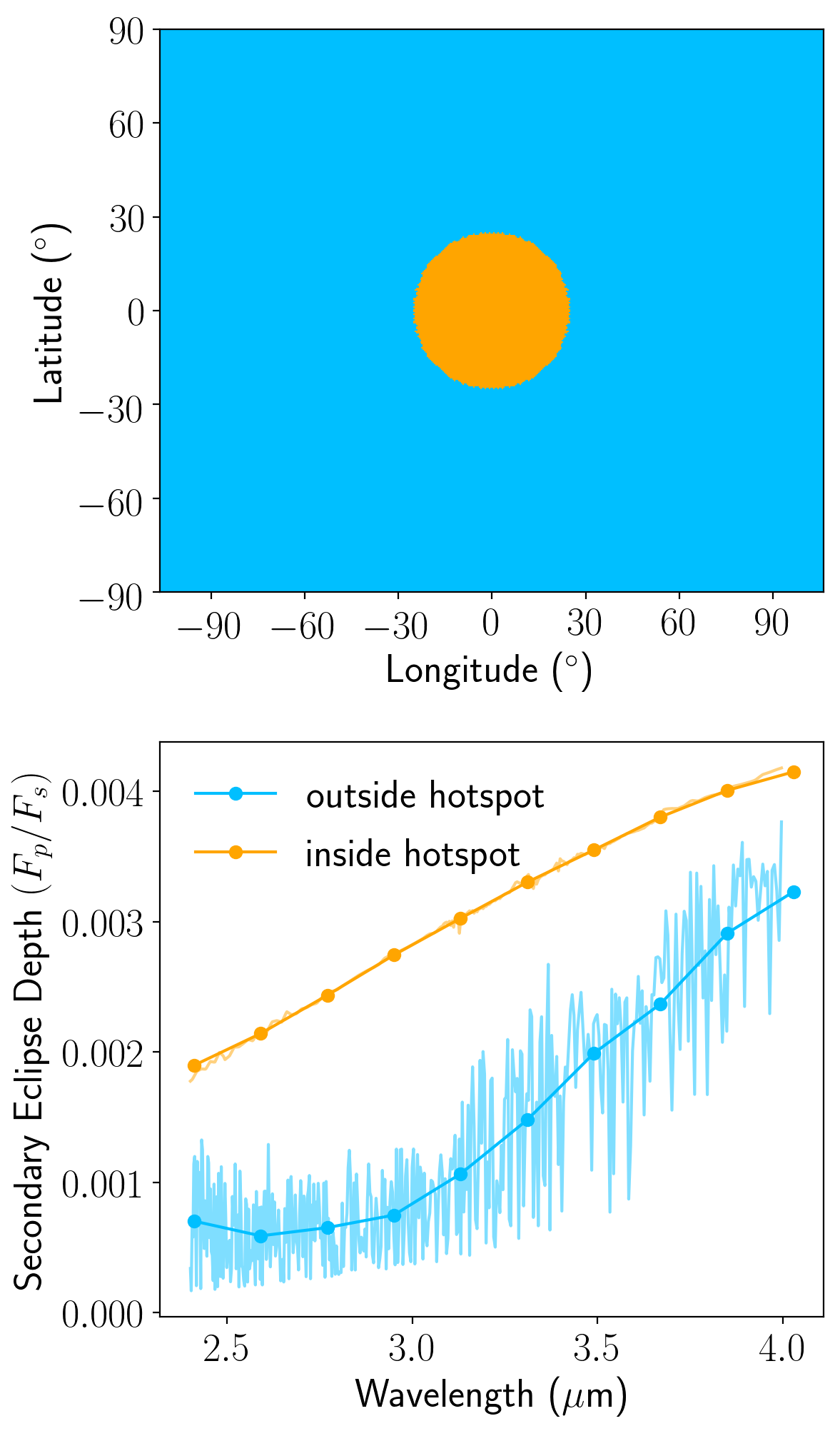}
    \caption{Input map (upper panel) and spectra (lower panel) for the Simplified Hotspot map. Colors indicate regions of the map where different spectra are painted on. Thin lines indicate unbinned spectra, and thick lines with points show spectra binned to the 10 wavelength bins used in our analysis. The map is centered on the substellar point and has a longitudinal and latitudinal extent set by the portion of the planet that is visible during our simulated secondary eclipse observing window.}
    \label{fig:hotspotinput}
\end{figure}

We demonstrate our method of identifying eigenspectra using three hypothetical maps constructed using the HEALpix projection \citep{Gorski2005healpix} for which we try to recover the input parameters with our methodology. The model planet-star system for all three maps is based on HD 189733b properties determined by \citet{stassun2017gaiaRadiiMasses}. 
Figure \ref{fig:hotspotinput} shows the first map, which we refer to as the ``Simplified Hotspot'' map, and the spectra used to construct it. 
The Simplified Hotspot map consists of one higher-flux spectrum painted within a circular region surrounding the substellar point, and a second lower-flux spectrum painted onto the rest of the planet. This mimics a potential, albeit simplified, eclipse map that could result from observation of a hot, synchronously rotating planet such as those that \textit{JWST} will observe. The hotspot has an angular diameter of $50\degr$. 

The spectra for the Simplified Hotspot are generated using the radiative transfer model described in detail by \citet{line2013chimera}. We deliberately choose atmospheric parameters that create two distinct spectra. The spectrum assigned to the area inside the hotspot only has methane (CH$_4$) as a molecular opacity source at a mixing ratio of $10^{-2}$ ppm. Outside of the hotspot, we input water (H$_2$O) as a sole molecular opacity source with a mixing ratio of $10^{3}$ ppm. We implement the same pressure-temperature profile parameterization to create both spectra: $\log{\gamma_1}$ = $-1$, $\log{\gamma_2}$ = $-1$, $\log{\kappa_{\rm IR}}$ = $-1$, $\alpha$ = $0.5$, $\beta$ = $1$. We refer the reader to \citet{line2013chimera} for a thorough description of these terms. Briefly, the profile follows one upwelling channel of thermal emission and two downwelling streams of visible radiation. The terms $\gamma_1$ and $\gamma_2$ correspond to the ratio of the Planck mean opacities of each visible stream to the thermal stream while $\alpha$ divides the flux between the two downwelling visible streams. The parameter $\beta$ describes the irradiation temperature based on stellar properties. The values we have chosen ensure that there is no thermal inversion in the profile.

\begin{figure}
    \centering
    \includegraphics[width=\linewidth]{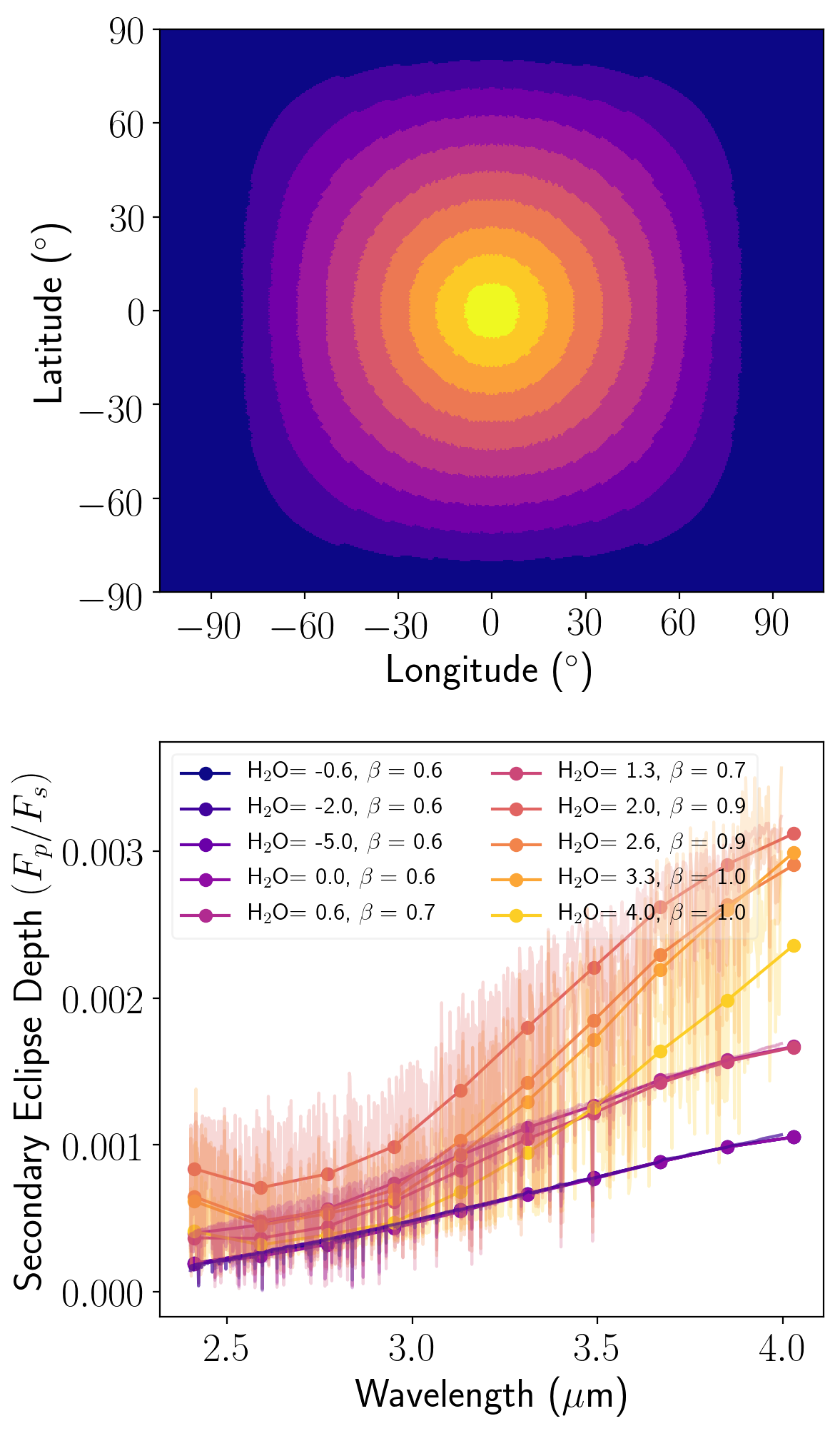}
    \caption{Input map (upper panel) and spectra (lower panel) for the continuum hotspot map. Colors indicate regions of the map where different spectra are painted on. Thin lines indicate unbinned spectra, and thick lines with points show spectra binned to the 10 wavelength bins used in our analysis. The map is centered on the substellar point and has concentric annular rings along which the fraction of incident flux that is absorbed ($\beta$) and log of the H$_2$O mixing ratio (in ppm) decrease radially from the substellar point.}
    \label{fig:conthotspotinput}
\end{figure}

The second map we construct is based on the idea that a realistic planet will likely not have two fully separable regions with different spectra, but rather will show a continuum of spectra between regions. Therefore, we construct 
the ``Continuum Hotspot'' map shown in Figure \ref{fig:conthotspotinput}, which has ten nested regions surrounding the substellar point. Each region has an angular width of $9\degr$ and is painted with a spectrum with a different water abundance and temperature profile, such that the spectra form a gradient of water abundance and temperature moving outward from the substellar point. 
We vary H$_2$O mixing ratio incrementally from $10^{-5}$ ppm to $10^{4}$ ppm and pair each input abundance with a temperature parameter selected from a range of $\beta$ = $0.6$ and $\beta$ = $1.0$. \citet{line2013chimera} define $\beta$ to encapsulate albedo, emissivity, and day-night redistribution, which determines the irradiation experienced by the planet from the host star; lower values correspond to cooler temperatures. We design the hemisphere such that the central hot spot is high in temperature and saturated with water; as we move to areas further from the hot spot, both the temperature and water abundance decrease. This gradient is not necessarily meant to represent what we think would occur on the dayside of a realistic planet, but is instead intended as a toy model to test the ability of our method to resolve gradual changes in temperature and chemistry across the hemisphere. 

\begin{figure}
    \centering
    \includegraphics[width=\linewidth]{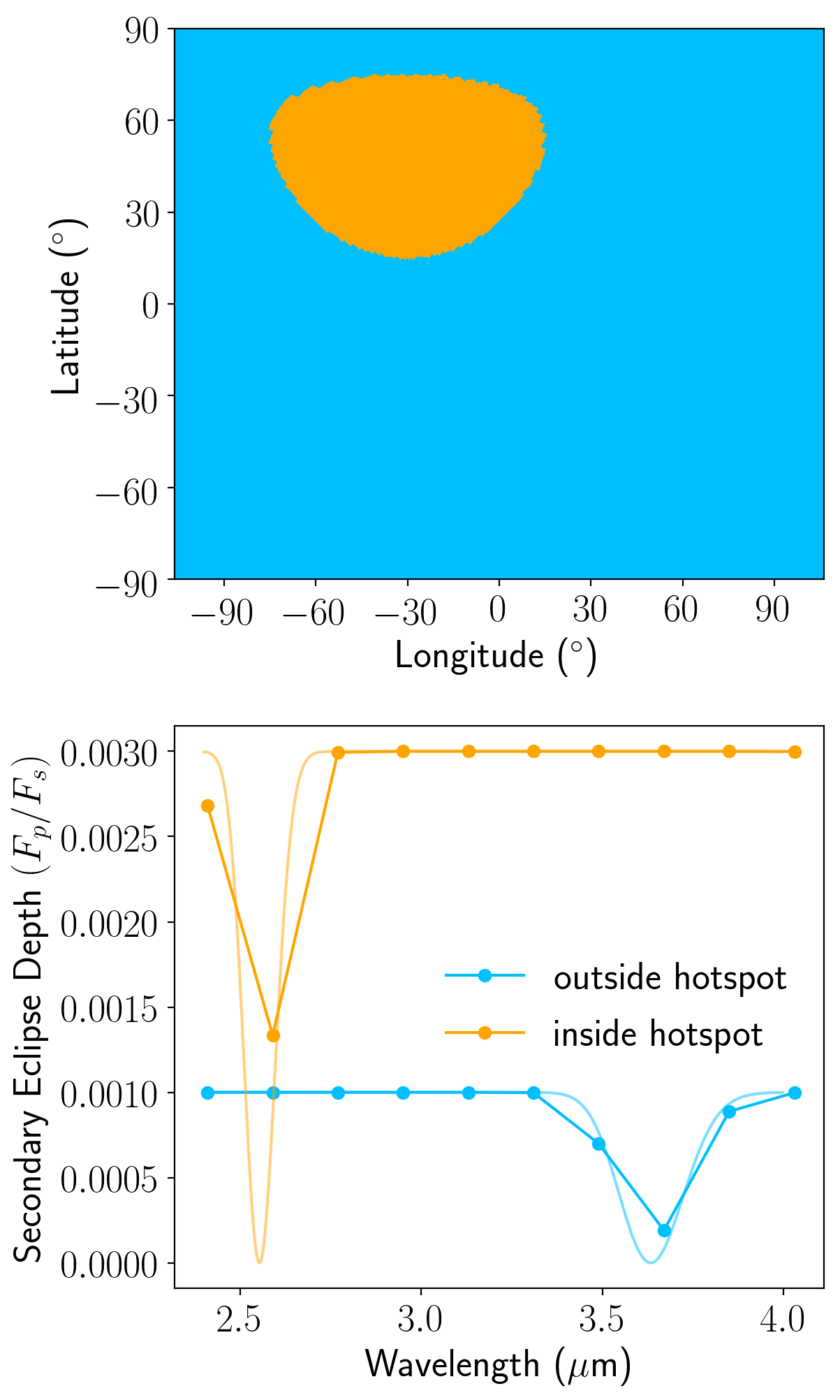}
    \caption{Input map (upper panel) and spectra (lower panel) for the asymmetric hotspot map. Colors indicate regions of the map where different spectra are painted on. Thin lines indicate unbinned spectra, and thick lines with points show spectra binned to the 10 wavelength bins used in our analysis. The map is centered on the substellar point and has a hotspot centered $45^{\circ}$ north and $30^{\circ}$ west of the substellar point, and spanning $60^{\circ}$ in angular diameter.}
    \label{fig:asymmhotspotinput}
\end{figure}

The third map (Figure~\ref{fig:asymmhotspotinput}), which we refer to as the ``Asymmetric Hotspot'' map, tests the ability of our spherical harmonic-based mapping method to constrain an asymmetric map with a shifted hot spot. It is similar to the Simplified Hotspot map in that it uses two easily-separable spectra, but the hotspot is offset from the substellar point, centered at $+45\degr$~latitude and $-30\degr$~longitude, and has an angular diameter of $60\degr$. For this test case, the two spectral components are designed to be easily separable and therefore differ in continuum flux, wavelength of their single spectral feature, and relative depth of that feature below the continuum. We specifically design the spectra in this map to have unrealistic shapes so we can separate out our ability to resolve structure across spatial dimensions on the map from our ability to resolve similarly-shaped spectra.

While we construct planet maps that cover the entire planet, we only consider observations during secondary eclipse and so can only constrain the planet's map on the dayside and the small fraction of the nightside we observe just before and after eclipse. However, the method we describe here could be used for full phase curve observations to produce a map covering the whole planet, although outside of secondary eclipse the observations would only be sensitive to variations with longitude, not latitude.

We use the analytic occultation code \texttt{starry}\footnote{We used \texttt{starry} version 0.3.0, now available at \url{https://github.com/rodluger/starry/tree/v0.3.0}.} \citep{luger2019starry} to model secondary eclipse light curves. We bin the high-resolution spectra at each HEALpix pixel to a lower resolution wavelength grid between 2.40 $\mu$m and 4.0 $\mu$m with a fixed $\Delta \lambda$ = $0.18$ $\mu$m, applicable to wavelength-binned data from the \textit{JWST}/NIRCam instrument using the F322W2 filter \citep{greene2017jatisNIRCam}. For each low-resolution wavelength interval, we expand the HEALpix map in spherical harmonics up to degree $l$ = $18$, and define a \texttt{starry.Secondary} planet object with this wavelength-dependent map. We note that the transformation from HEALpix to spherical harmonics introduces a small amount of error into both the map and the spectra, but this amount is well below the precision of our simulated observations. Finally, we use \texttt{starry} to compute analytic secondary eclipse light curves at each wavelength. This results in 10 light curves, which are each normalized to the out-of-eclipse continuum flux at the respective wavelength. A vertical shift is applied so the bottom of the eclipse is defined to be zero planetary flux (but still 100\% stellar flux). 

We simulate a \textit{JWST} time series observation corresponding to the \texttt{starry} light curves. 
As in \citet{schlawin2018JWSTforecasts}, we use the \texttt{pynrc}\footnote{\url{https://pynrc.readthedocs.io/en/latest/}} NIRCam observation simulator to calculate the signal to noise of the spectrum.
The signal to noise per integration is used to create a time series for each wavelength.

When creating the simulated time series, we add error bars based on the NIRCam observation simulation but do not  
actually add random noise to the time series.

\subsection{Extracting the Eigenspectra from Simulated Observations}
\label{sec:extract}

\begin{figure*}
    \centering
    \includegraphics[width=\linewidth]{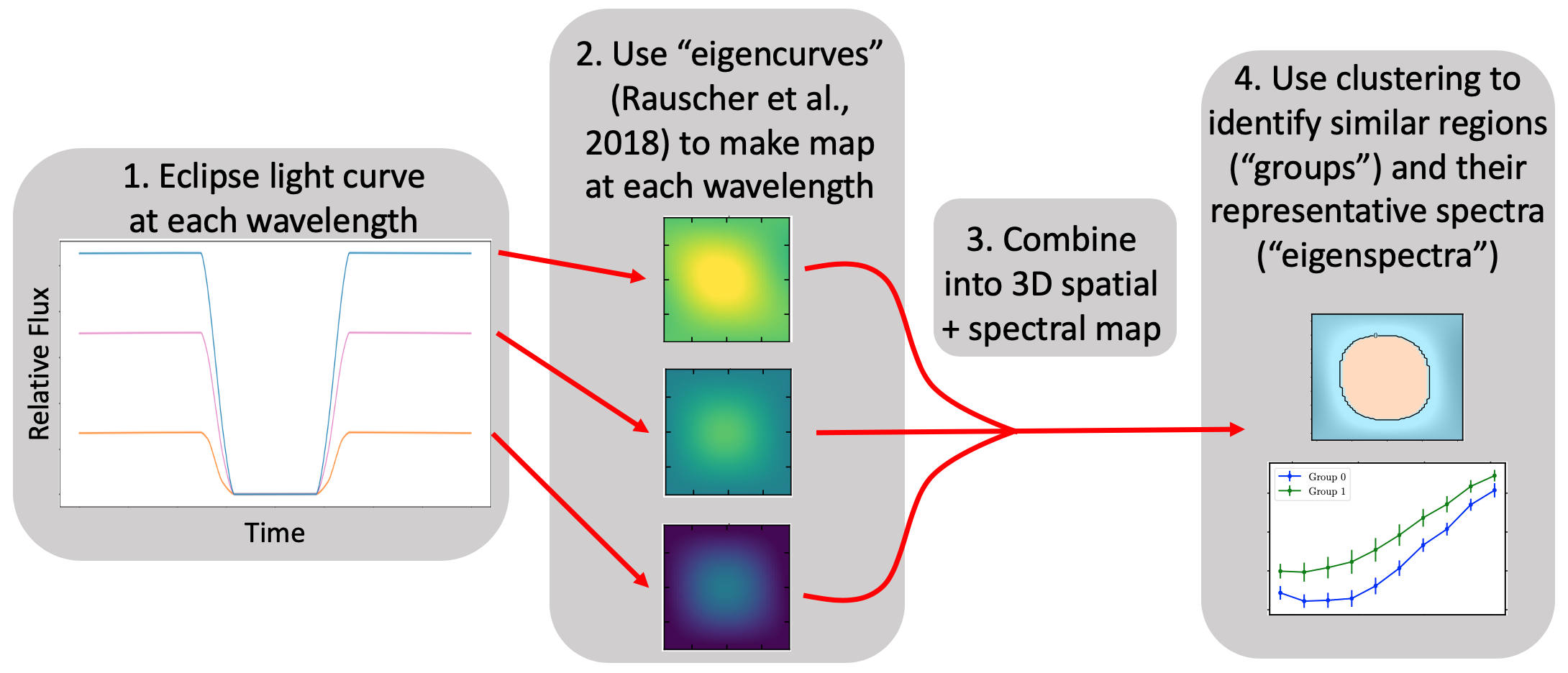}
    \caption{Overview of the process we use to extract eigenspectra from the eclipse light curves. We apply the method of \citet{Rauscher2018} and use eigencurves to construct a map separately at each wavelength. We then combine these single-wavelength maps into a 3D spatial + spectral map. We use K-means clustering to identify similar regions on this 3D map (``groups'') and their representative spectra (``eigenspectra'').}
    \label{fig:cartoon}
\end{figure*}

\begin{figure}
    \centering
    \includegraphics[width=\linewidth]{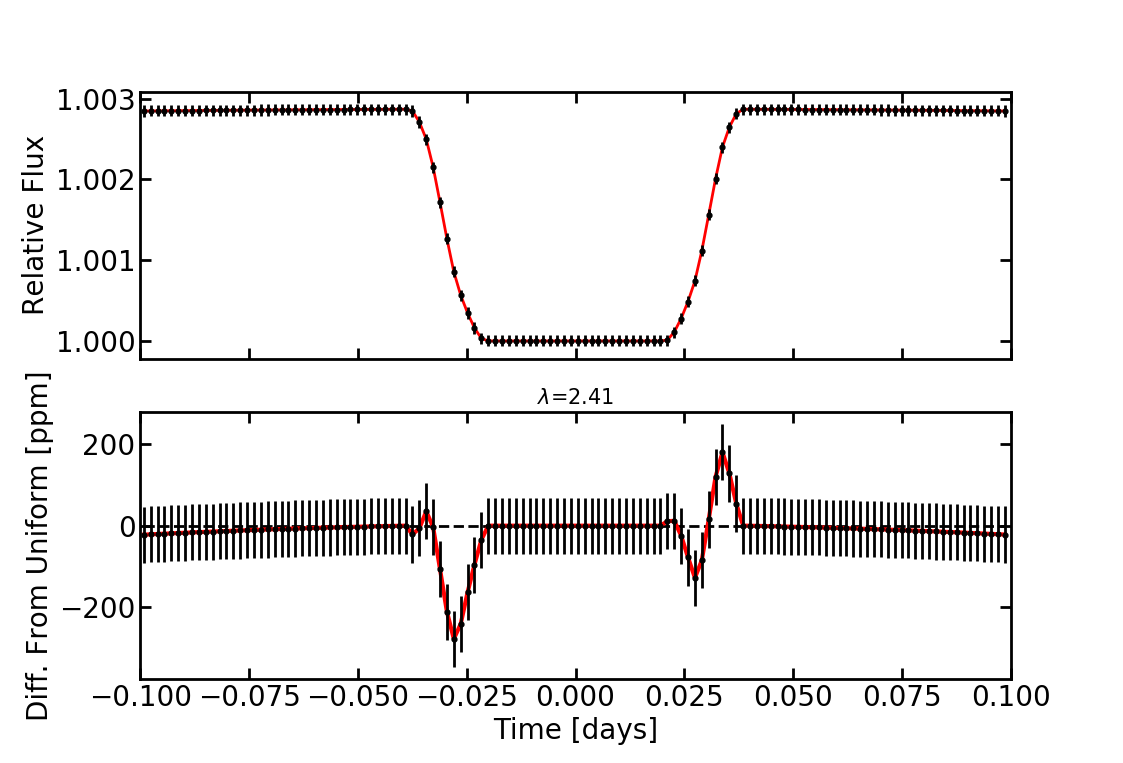}
    \caption{Top: Example light curve showing simulated data (black points) at $\lambda$ = $2.41$~$\mu$m for the Simplified Hotspot model. Red line shows the fit using the eigencurve method of \citet{Rauscher2018}. Bottom: Difference in flux from a uniform sphere for our eigencurve fit (red line) and simulated data (black points).}
    \label{fig:lightcurve}
\end{figure}

Figure~\ref{fig:cartoon} shows an overview of the process we use to extract the eigenspectra. We create a set of light curves from 
spherical harmonics using the \texttt{spiderman} package \citep{louden2017spiderman}.
We use \texttt{spiderman} for this step and \texttt{starry} in Section \ref{sec:construct} because of existing legacy code and because using two separate codes for injection and recovery makes the process less circular.
We include spherical harmonics up to $l$ = $2$ because using this many harmonics provides 8 linearly independent eigencurves, many more than can be constrained by eclipse observations at the precision of our simulated measurements \citep{Rauscher2018}. However, we include in our code the capability to fit for higher-order spherical harmonics. We follow the methods of \citet{Rauscher2018} and use principal component analysis (PCA) to construct orthogonal light curves from linear combinations of the spherical harmonics. Using these orthogonal ``eigencurves'' instead of directly fitting for spherical harmonic coefficients reduces the correlations between parameters, and the PCA produces a list of eigencurves ranked by their relative potential contribution to the observed light curve. For each wavelength that we fit for, we select the number of eigencurves to use at that wavelength by determining the largest number of eigencurves before any of them show significant cross-correlation with each other (see Section~\ref{sec:numeigencurves} for a discussion of how correlated eigencurves result in a less accurate map). Figure~\ref{fig:lightcurve} shows simulated observations for a single wavelength of the Simplified Hotspot model and the resulting fit.

We estimate the contributions of each eigencurve with a Markov Chain Monte Carlo (MCMC) fit using the \texttt{emcee} package \citep{foreman-mackey2013emcee}. We use 100 walkers and a chain of 3000 steps with a 300-step burn-in. We test for convergence by computing the autocorrelation time for each free parameter and ensuring that the number of samples is at least 50 times larger than the autocorrelation time.
We construct an extracted planet map by calculating the contributions of each eigencurve to each coefficient of the spherical harmonics for the map. 
The MCMC routine returns several hundred realizations of each set of fit components, so we quantify the propagated uncertainty in the flux maps by calculating the mean and variance in the flux at each point across all realizations.

We use K-means clustering to identify regions of the retrieved brightness map with similar spectra. K-means clustering is an algorithm that groups a number of observations $n$, which can be vectors, into $K$ clusters \citep{pedregosa2011scikit-learn}. 
We use K-means clustering rather than PCA because K-means provides the capability to cluster in multidimensional space, so we can identify regions that are spectrally similar. PCA could be used to pick out orthogonal spectral features that have the largest variance across the map but not group the spectra into similar categories. 
A further advantage with K-means is that its output can be more easily turned into a map because it assigns each spectrum to a single group, whereas PCA would output what percentages of each spectrum come from each principal component.
We also note that the changes in spectral features due to a cloud or chemical difference may not necessarily be orthogonal.

We select 100 random samples from the MCMC chain to perform clustering on. For each sample, we divide the extracted planet map into $n$ = $10^{4}$ sectors (100 divisions each in latitude and longitude). We input the spectra from each section of each sample's map into the clustering algorithm. We treat the spectrum at each point as a multi-dimensional vector, and group the set of spectra into $K$ groups and 10 spectral bins. This allows us to identify regions on the retrieved map with similar spectra. From this, we take the mean of all the spectra in each group as the representative spectrum, or ``eigenspectrum,'' of that group. The errors on each group's ``eigenspectrum'' are the standard deviation of all of the spectra from all of the maps which were identified as belonging to that group. We note that correctly propagating errors through a non-deterministic method such as K-means clustering is not straightforward, so we leave a more detailed study of the correct error propagation for future work. As we show in Section~\ref{sec:results}, the method we use in this paper is sufficient to identify large-scale spatial and spectral features in our simulated observations.

\section{Mapping Results and Discussion}
\label{sec:results}

\begin{figure*}
\centering
\includegraphics[width=0.3\linewidth]{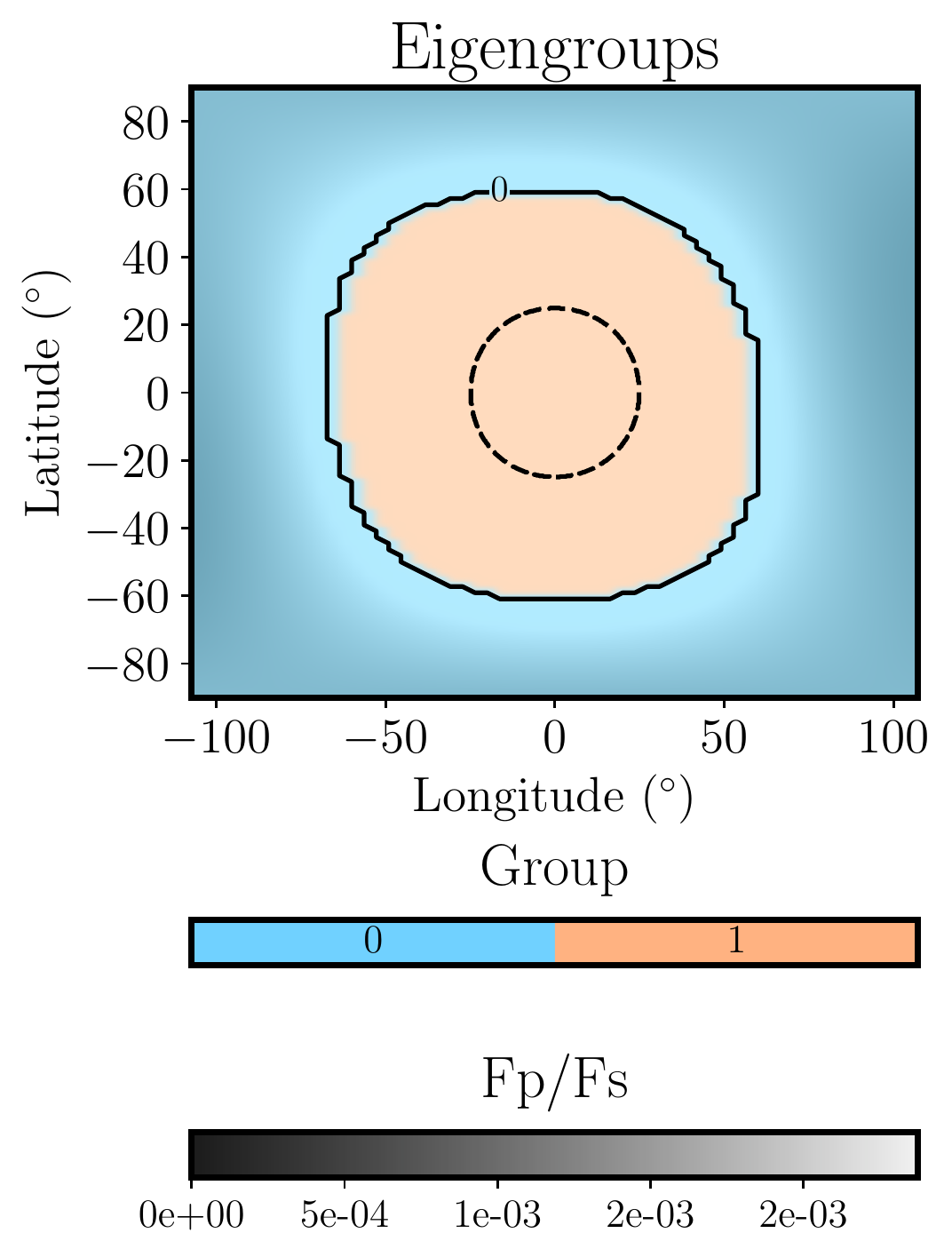}\hfill 
\includegraphics[width=0.3\linewidth]{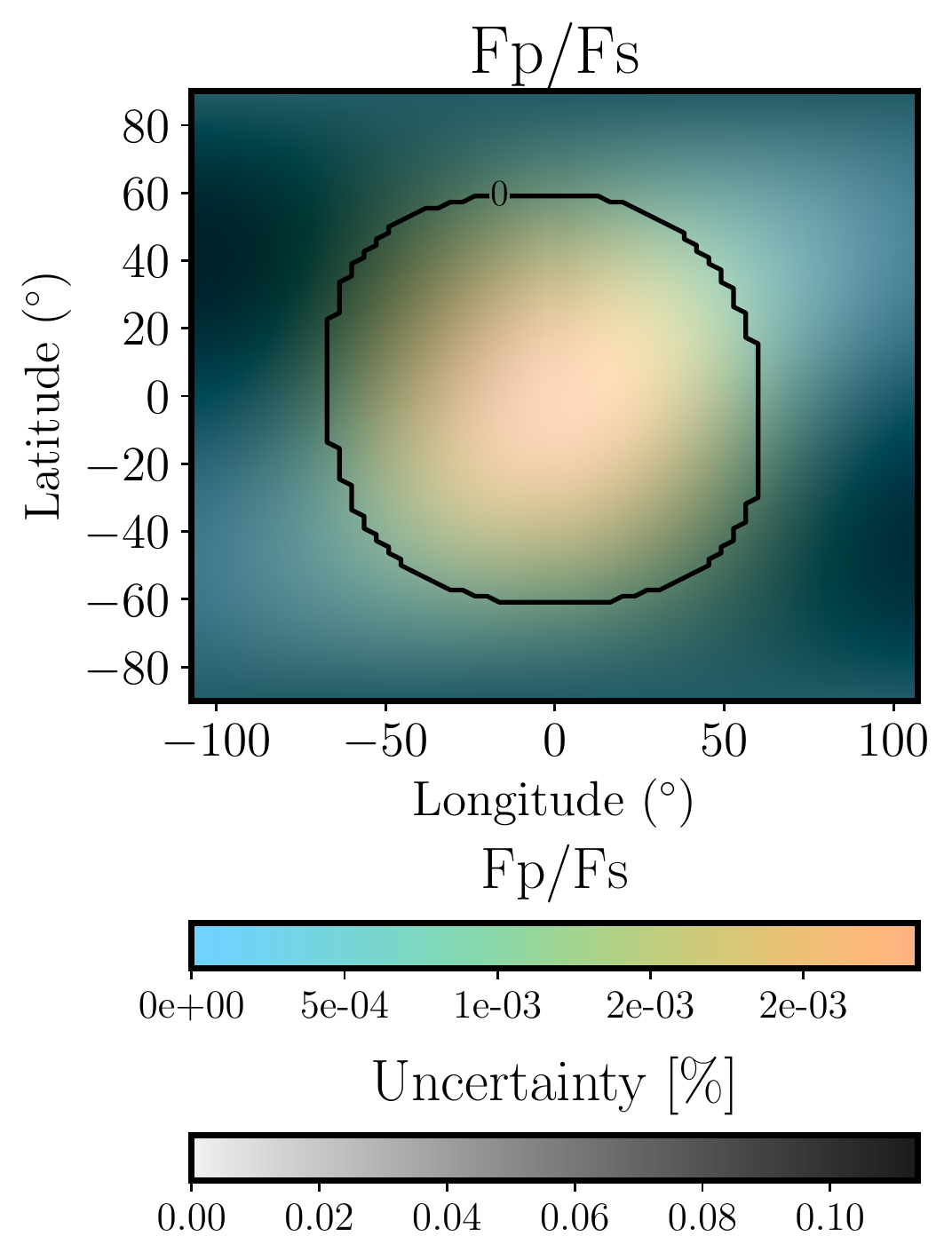}\hfill 
\includegraphics[width=0.3\linewidth]{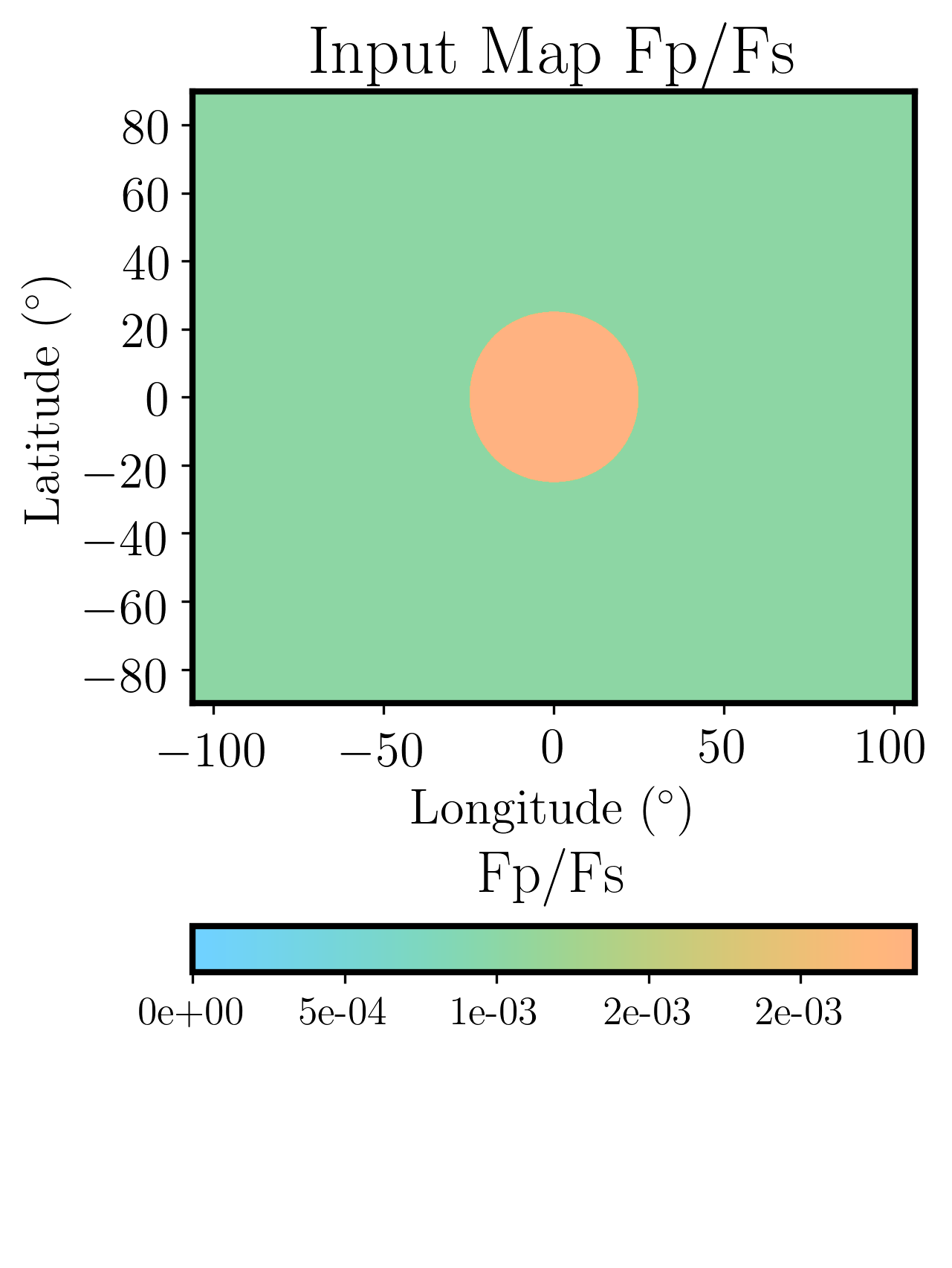}
\caption{Retrieved spectral group maps for the Simplified Hotspot model. The solid contours delineate separations by best-fit groups in each plot. \emph{Left:} The hue represents the best-fit group from the K-means clustering algorithm, and the brightness represents the mean intensity ratio ($F_{p}$/$F_{s}$) in the observed wavelength range. Note that here and in Figures~\ref{fig:ContinuumMap} and \ref{fig:OffsetMaps} we use $F_{p}$ to refer to the intensity of the planet at that point multiplied by the planet's solid angle, so that the ratio $F_{p}$/$F_{s}$ is unitless. The dashed contour indicates the extent of the hotspot in the original input model. \emph{Middle:} The hue represents the mean $F_{p}$/$F_{s}$, and the brightness represents the uncertainty in the eclipse depth. \emph{Right:} The mean $F_{p}$/$F_{s}$ from the input map, on the same color scale as the output map in the middle panel.}
\label{fig:SpectralGroupMap}
\end{figure*}

\begin{figure*}
\begin{center}
\begin{tabular}{cc}
\includegraphics[width=0.9\columnwidth]{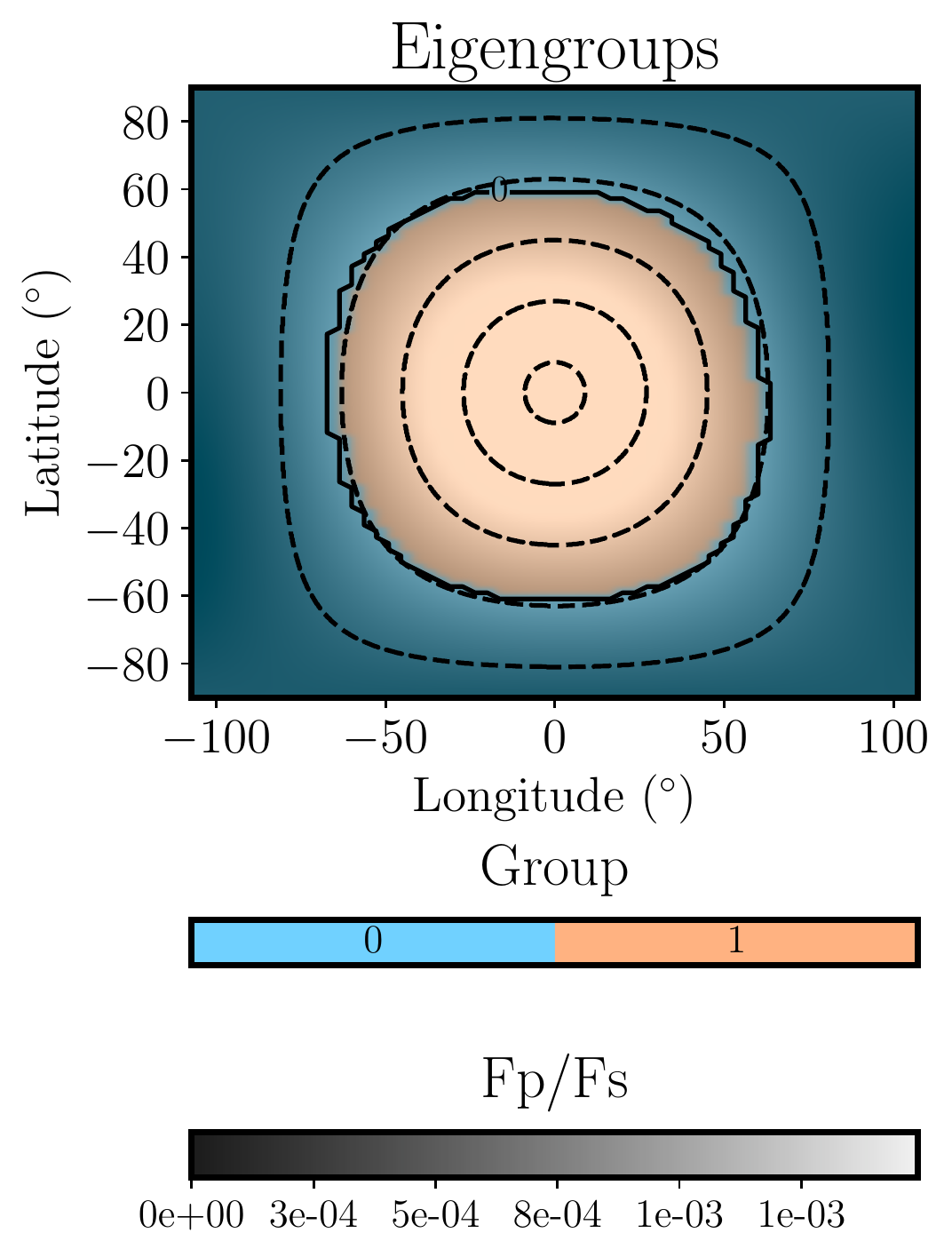} &
\includegraphics[width=0.9\columnwidth]{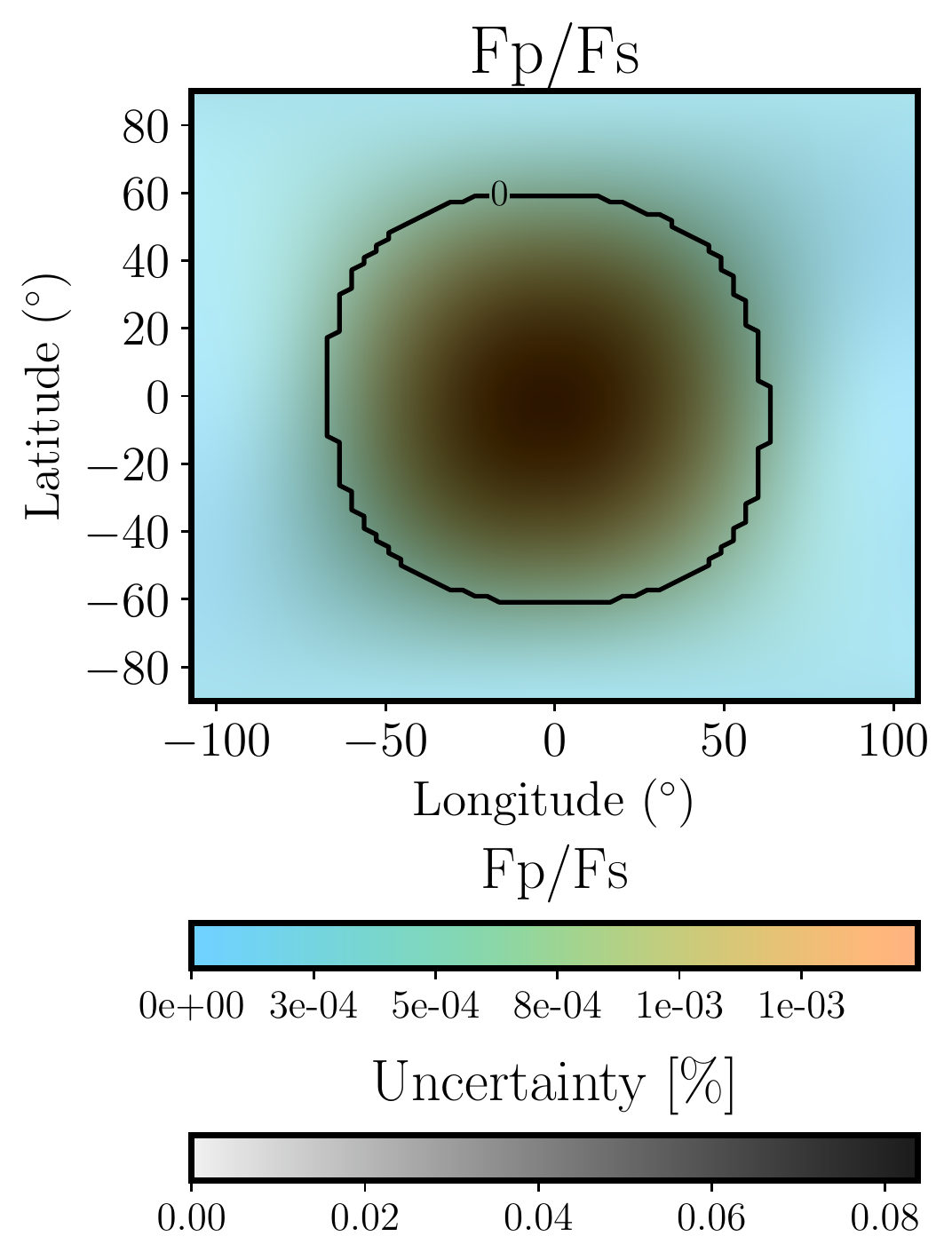} \\
\end{tabular}
\caption{Same as first two panels of Figure~\ref{fig:SpectralGroupMap}, but for the Continuum Hotspot model. For clarity, the dashed lines show the boundary of every other ring in the original input model.}
\label{fig:ContinuumMap}
\end{center}
\end{figure*}

\begin{figure*}
\begin{center}
\begin{tabular}{cc}
\includegraphics[width=0.9\columnwidth]{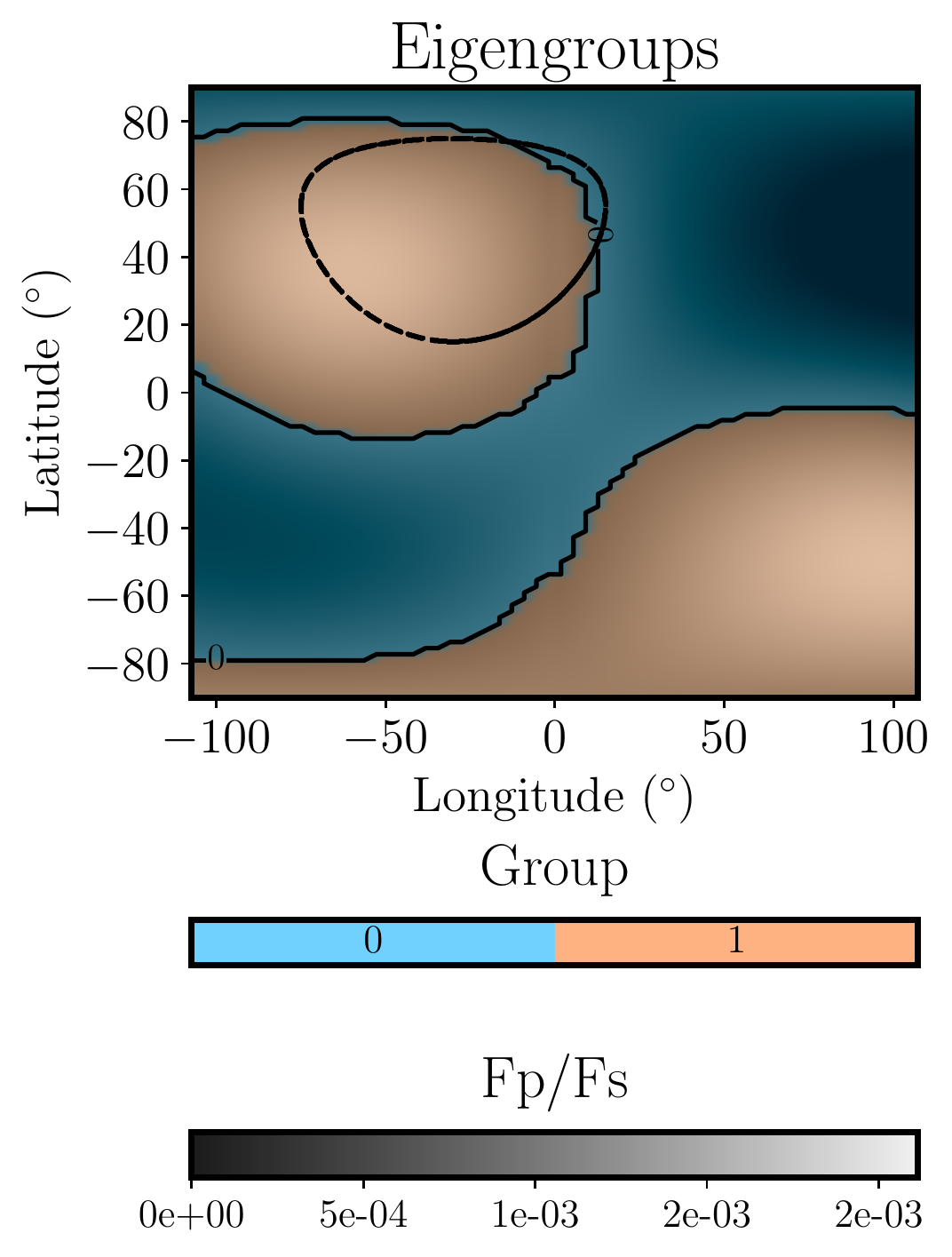} &
\includegraphics[width=0.9\columnwidth]{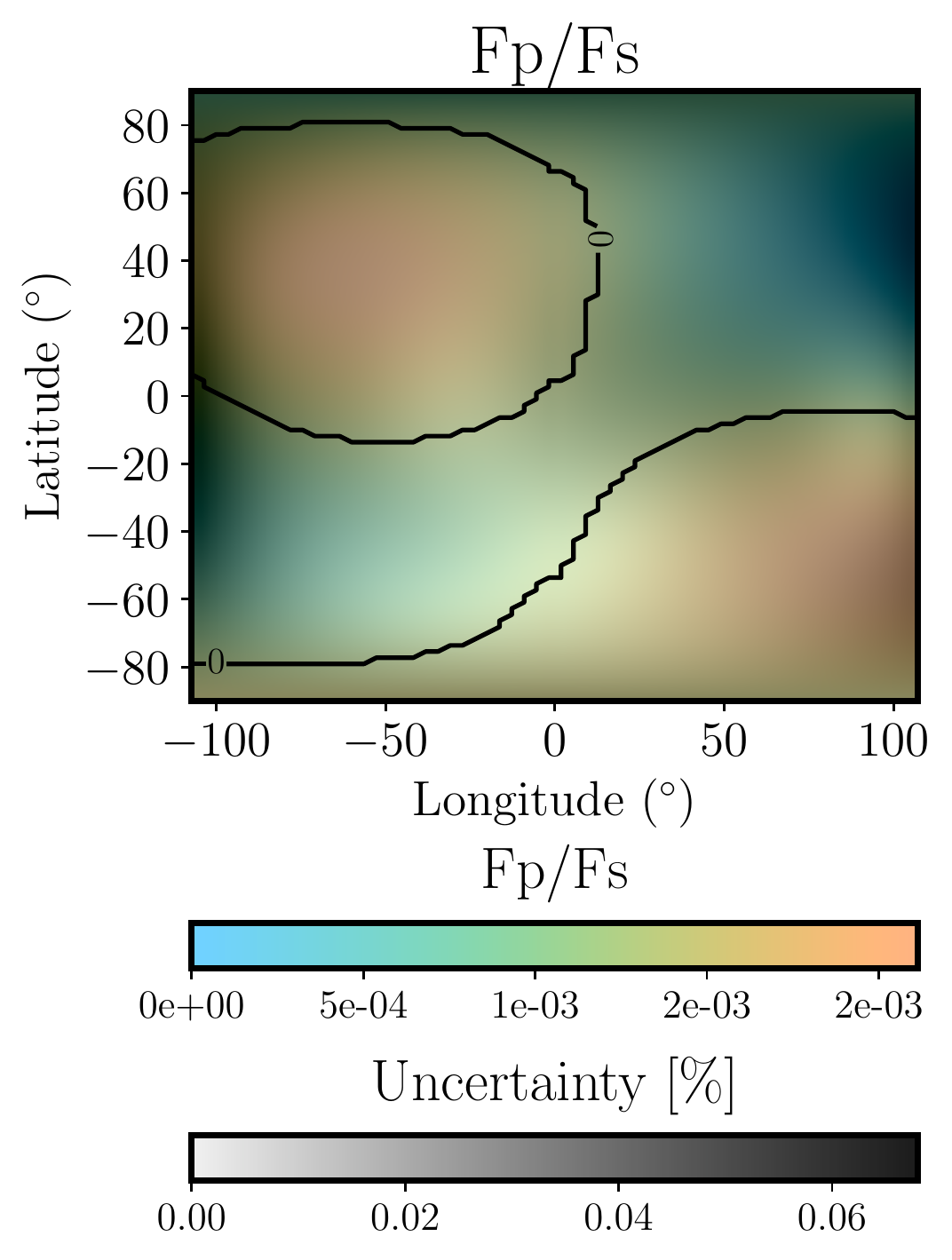} \\
\end{tabular}
\caption{Same as first two panels of Figure~\ref{fig:SpectralGroupMap}, but for the Asymmetric Hotspot model.}
\label{fig:OffsetMaps}
\end{center}
\end{figure*}

\begin{figure}
    \centering
    \includegraphics[width=\linewidth]{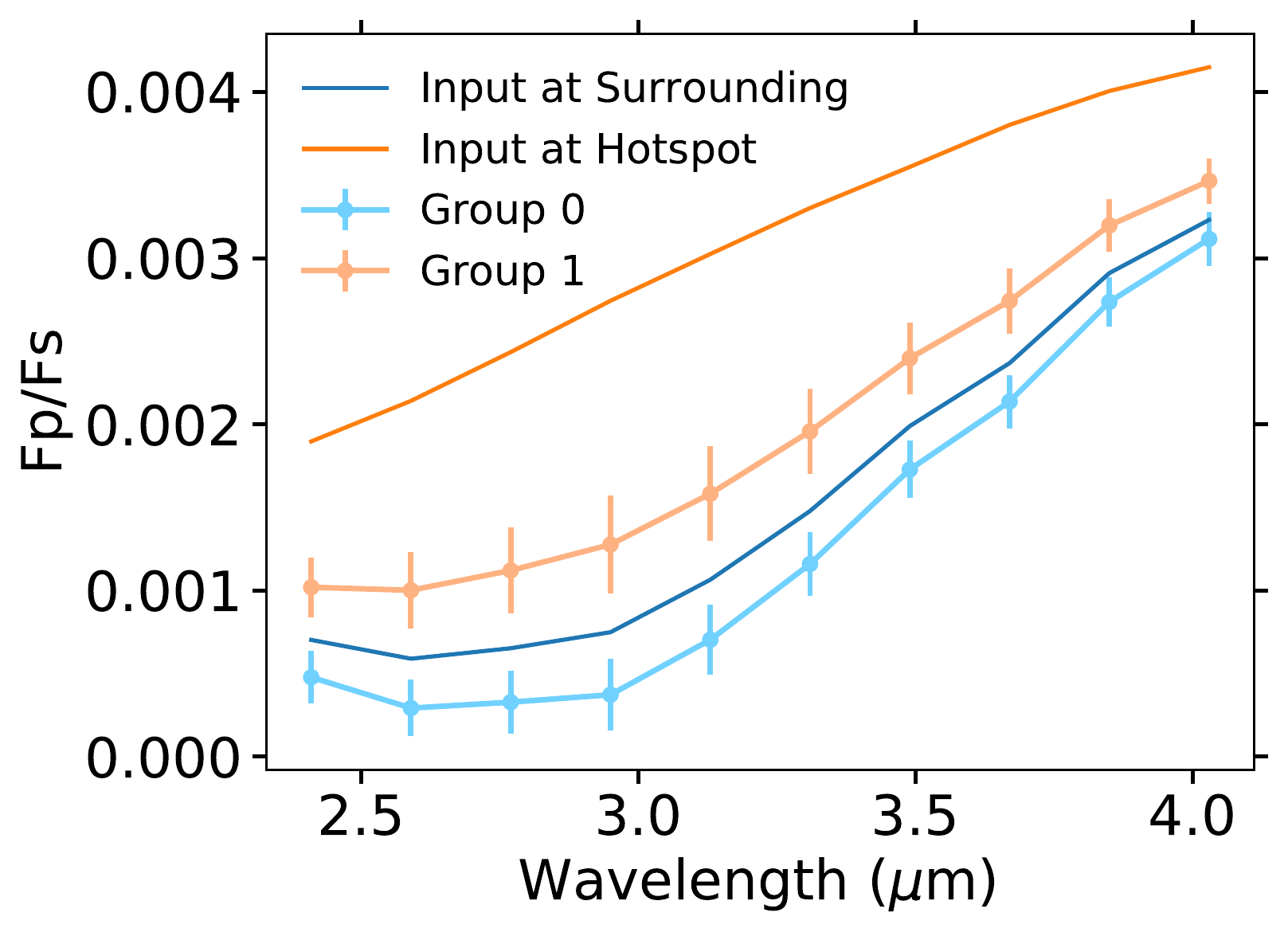}
    \caption{Results of K-means clustering to recover two groups from the Simplified Hotspot map. The orange and blue lines without error bars show the original input spectra from Figure~\ref{fig:hotspotinput}, while the lines with error bars show output spectra for the regions of the map assigned to Group~0 (surrounding the hotspot) and Group~1 (inside the hotspot). The K-means clustering method identifies a larger hotspot than the input map, which results in some mixing of the input spectra, but generally correctly identifies the spectral shapes.}
    \label{fig:outputspec}
\end{figure}

\begin{figure}
    \centering
    \includegraphics[width=\linewidth]{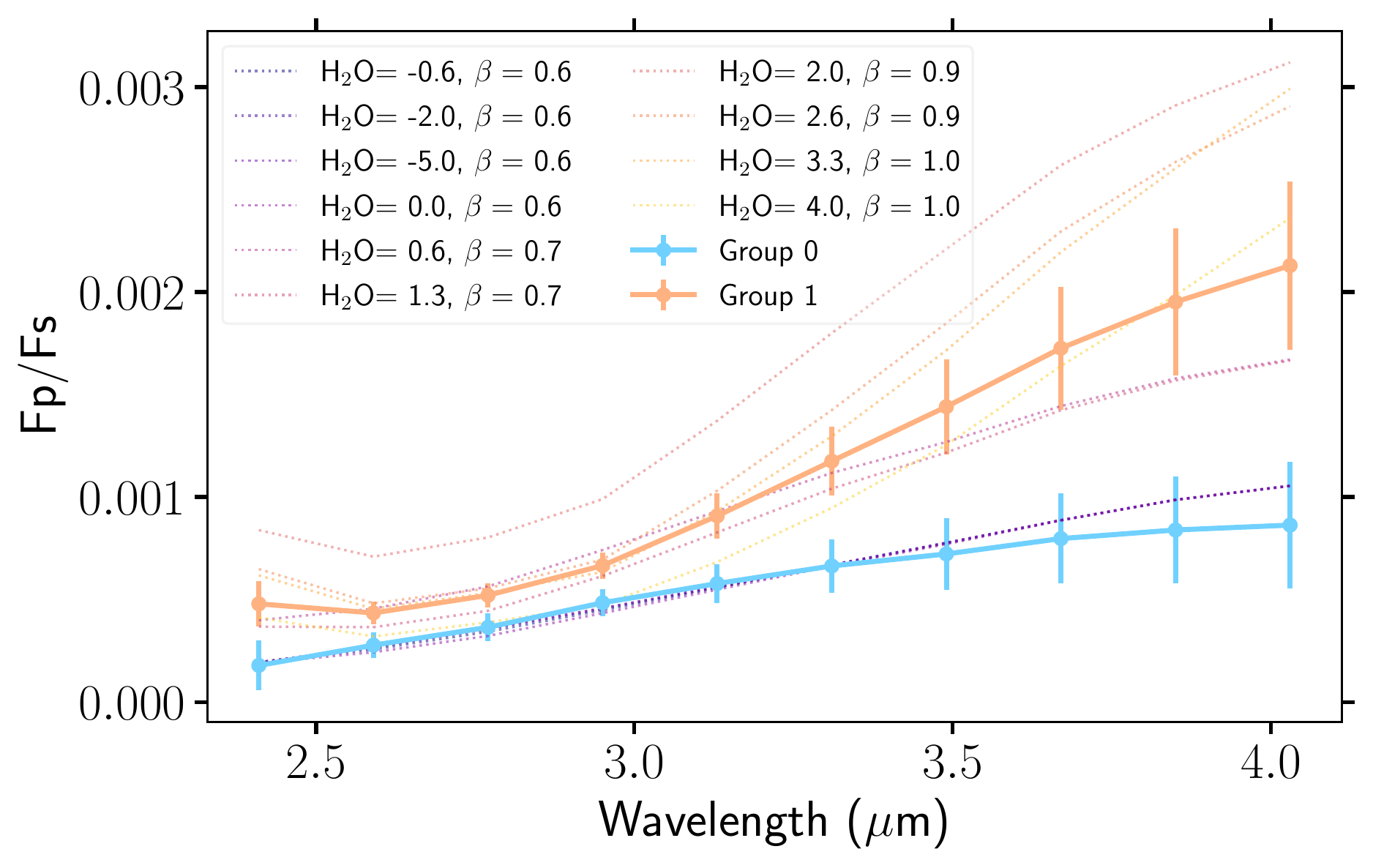}
    \caption{Results of K-means clustering to recover two groups from the Continuum Hotspot map, which has ten distinct input spectral groups. The dotted lines without error bars show the original input spectra from Figure~\ref{fig:conthotspotinput}, while the lines with error bars show output spectra for the regions of the map assigned to Group~0 (surrounding the hotspot) and Group~1 (inside the hotspot). The K-means clustering method delineates the groups at a radial distance intermediate between the center and boundary of the input continuum hotspot.}
    \label{fig:outputspec_continuum}
\end{figure}

\begin{figure}
    \centering
    \includegraphics[width=\linewidth]{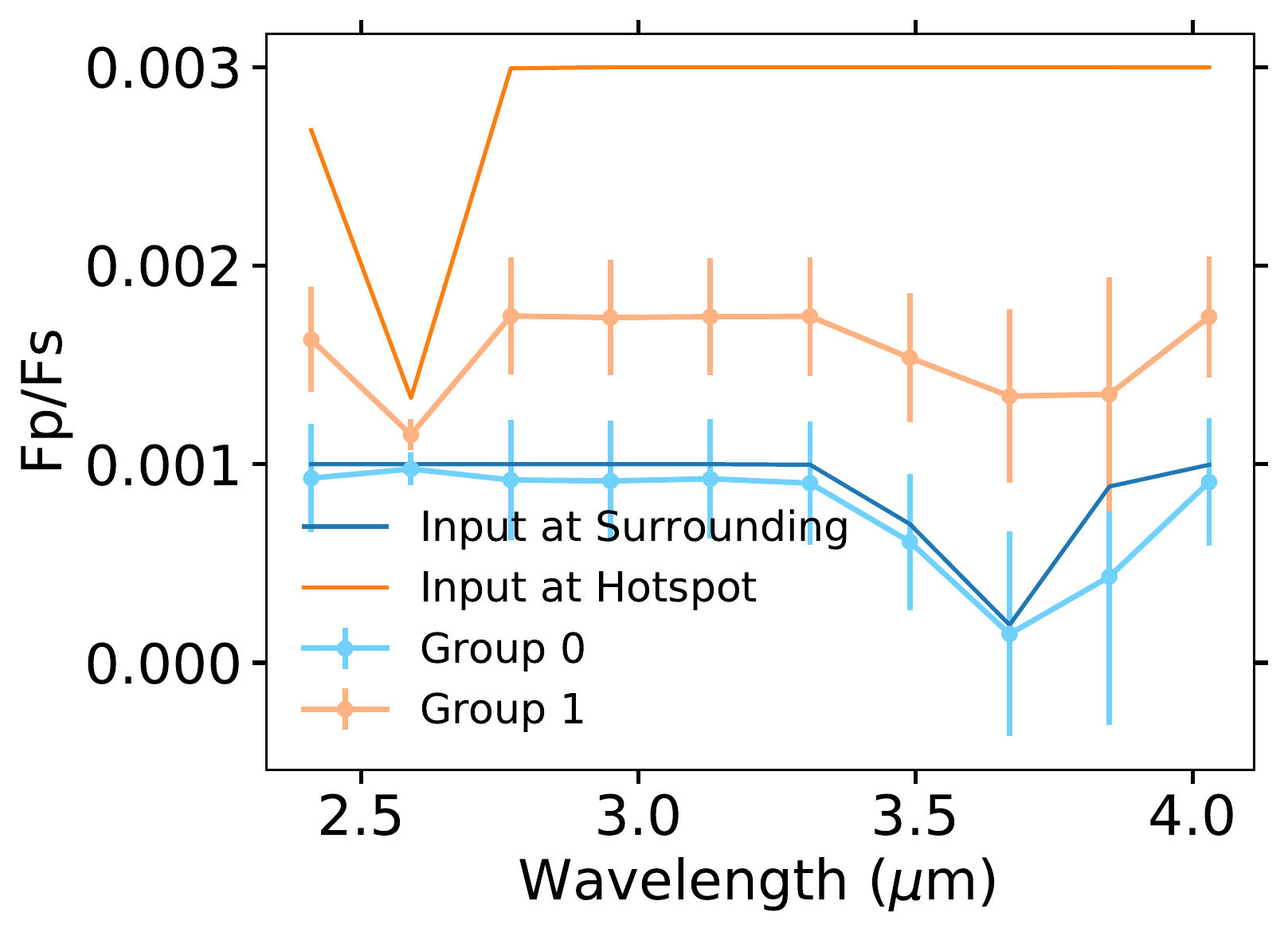}
    \caption{Results of K-means clustering to recover two groups from the Asymmetric Hotspot map. The orange and blue lines without error bars show the original input spectra from Figure~\ref{fig:hotspotinput}, while the lines with error bars show output spectra for the regions of the map assigned to Group~0 (surrounding the hotspot) and Group~1 (inside the hotspot). The K-means clustering method identifies a secondary hotspot, which results in some mixing of the input spectra, but generally correctly identifies the spectral shapes.}
    \label{fig:outputspec_asymmetric}
\end{figure}

Figures~\ref{fig:SpectralGroupMap}, \ref{fig:ContinuumMap}, and \ref{fig:OffsetMaps} show the output flux maps and groupings our pipeline produces for the Simplified Hotspot, Continuum Hotspot, and Asymmetric Hotspot maps, respectively. Figures~\ref{fig:outputspec}, \ref{fig:outputspec_continuum}, and \ref{fig:outputspec_asymmetric} show the corresponding eigenspectra retrieved from these maps compared to the input spectra. In Section~\ref{sec:hotspot} we use the Simplified Hotspot map to discuss how our pipeline performs in an idealized case. We use the Continuum Hotspot model to test the limits of our pipeline's ability to retrieve spectral information as quantified by the number of significant eigencurves (Section~\ref{sec:numeigencurves}) and groups (Section~\ref{sec:numgroups}). Finally, in Section~\ref{sec:asymmetric}, we use the Asymmetric Hotspot map to test how well our pipeline can represent flux distributions that aren't symmetric about the substellar point.

\subsection{The Simplified Hotspot Map}
\label{sec:hotspot}

The Simplified Hotspot map was originally created using two distinct spectra, so to test the ability of our methods to recover these spectra we use K-means clustering to create two groups with different eigenspectra. 
Figure \ref{fig:SpectralGroupMap} shows two ways to visualize the eigenspectra groupings on the Simplified Hotspot model. This figure shows the areas of the map grouped into the two eigenspectra, the calculated fluxes, and the uncertainties in the fluxes at a wavelength of 2.43 $\mu$m. The rightmost plot in this Figure shows the original input map on the same flux scale as the output flux map for comparison. Figure~\ref{fig:SpectralGroupMap} shows that the K-means clustering generally identified the structure of our input map correctly, although comparison to Figure~\ref{fig:hotspotinput} shows that the hotspot output by the K-means clustering is more spatially extended than the input hotspot. Our maps are constructed using the eigencurves method of \citet{Rauscher2018}, which can not reproduce sudden discontinuities in flux across the map, so our inability to identify the exact extent of the hotspot is likely because the eigencurves can not perfectly represent the sharp edge between the two groups. We construct this map using only the first four eigencurves because adding more eigencurves would result in larger error bars (see Section~\ref{sec:numeigencurves}). This means our map is limited to large-scale flux differences and would not be able to show small-scale changes \citep{Rauscher2018}, which may be another reason for the broadening of the hotspot in our output map compared to the input map. Our inability to identify small-scale flux changes when representing the map with just a few eigencurves explains why the output map shows a high precision even in regions where the output map flux is many sigma away from the input map flux (for example, at around 30$\degr$ away from the substellar point, where the input map shows a low flux outside the hotspot but the output map precisely identifies a high-flux hotspot region). Each realization of our output map from the MCMC correctly identifies that there is a substellar hotspot, but the angular size of this hotspot is restricted by the small number of eigencurves we use. Therefore, our output map shows a small uncertainty in this region because each realization of the map shows a very similar flux distribution there, despite the fact that this flux distribution does not match the ``true'' input distribution. We discuss in Section~\ref{sec:numeigencurves} why we restrict ourselves to this small number of eigencurves in this paper and methods for incorporating information from larger numbers of eigencurves, which could potentially identify smaller-scale features.

This slight mixing of the areas on the edges of the two input groups can also be seen in Figure~\ref{fig:outputspec}, which shows the eigenspectra for the two groups identified by the K-means clustering compared to the input spectra for the hotspot and surrounding area. The K-means clustering correctly identifies a higher-flux central region and a lower-flux surrounding region. However, the clustering algorithm includes some of the surrounding area in the hotspot, which dilutes it and leads to a lower-flux spectrum for the output Group~1. Despite this dilution, our method correctly identifies the general shape of the input map and spectra.

\subsection{How many eigencurves should be used?}
\label{sec:numeigencurves}

We tested modeling the Continuum Hotspot map with different numbers of eigencurves to determine a best practice for selecting how many eigencurves to include when analyzing a set of data. 
\citet{Rauscher2018} discuss how adding more eigencurves to a fit eventually results in maps that are more uncertain than those with fewer eigencurves. We find a similar result using the Continuum Hotspot map. We test modeling this map with up to five eigencurves and find that four eigencurves provide a good fit to the synthetic data, while at all wavelengths using five eigencurves introduces degeneracies that make the results more uncertain and increases the variance in the fits.

\begin{figure}
    \centering
    \begin{subfigure}{}
    \includegraphics[width=\linewidth]{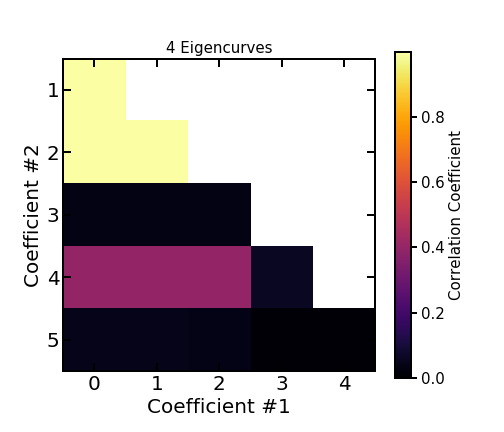}
    \end{subfigure}
    \vspace{-5 mm}
    \begin{subfigure}{}
    \includegraphics[width=\linewidth]{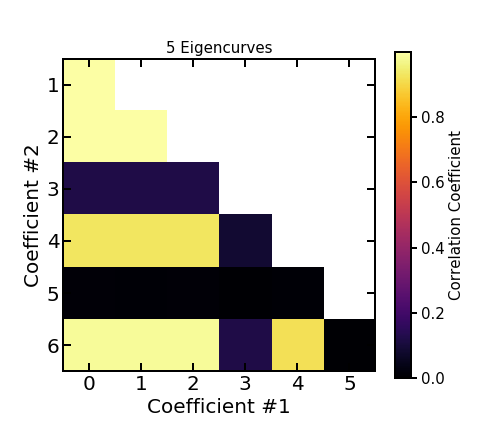}
    \end{subfigure}
    \caption{Diagrams showing cross-correlation coefficients between the eigencurves at a wavelength of 2.41~$\mu$m for models of the Continuum Hotspot using four and five eigencurves (top and bottom plots, respectively). For the case with four eigencurves, the only coefficients that are strongly correlated with any of the others are the first two coefficients, which represent the uniform-planet-brightness coefficient and a correction to the stellar flux. As described in \citet{Rauscher2018}, these two coefficients are not orthogonal by design the way the rest of the eigencurve coefficients are, so they are expected to show some correlation with the other coefficients. However, with five eigencurves there are several other eigencurve coefficients which are significantly correlated with each other. This indicates that the model produces a degenerate solution when more eigencurves are included than the amount that can be well-constrained at the noise level of the data \citep{Rauscher2018}.}
    \label{fig:crosscorr}
\end{figure}

\begin{figure}
    \centering
    \includegraphics[width=\linewidth]{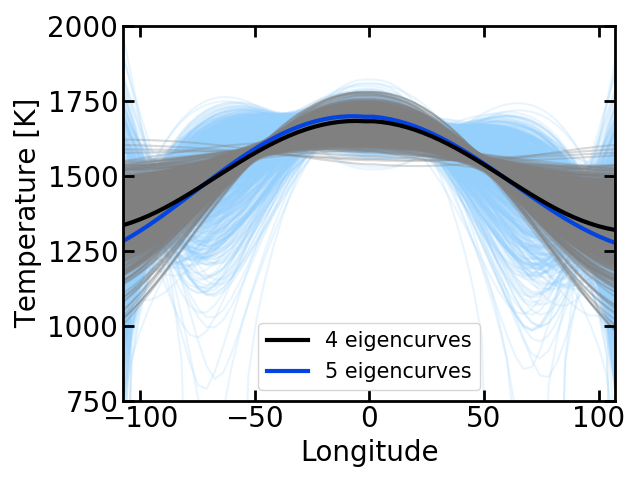}
    \caption{Brightness temperature along the equator as a function of longitude for the Continuum Hotspot Map, modeled using four (black) and five (blue) eigencurves. The thick line shows the best-fit solution, while the thin lines show 1000 random samples from the MCMC chain for each fit. Both models converge on similar best-fit solutions. However, the model using four eigencurves, which can be well constrained at the noise level of the data, shows a smaller spread in the derived temperatures. The model using five eigencurves shows a larger spread in the derived temperatures, including some MCMC samples where the temperature drops to unphysical negative values at some longitudes, because this model contains more eigencurves than can be well constrained by the data and is instead driven by the uninformative prior on the fifth eigencurve \citep{Rauscher2018}.}
    \label{fig:tempvslong}
\end{figure}

\begin{figure}
    \centering
    \includegraphics[width=\linewidth]{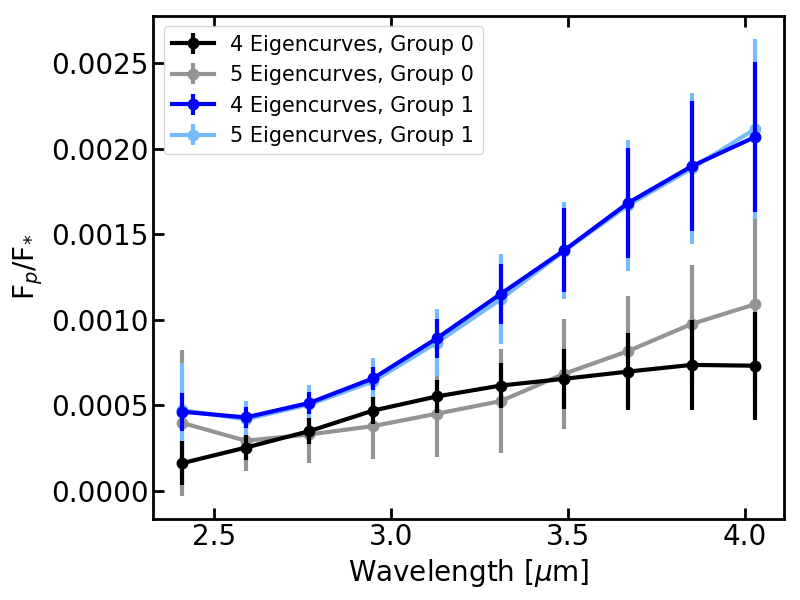}
    \caption{Eigenspectra for the Continuum Hotspot model when using four (dark blue and black points) vs. five (light blue and grey points) eigencurves. A model using five eigencurves results in spectra with larger error bars because some of the eigencurves are correlated, which leads to increased uncertainty in the planet map.}
    \label{fig:changeneigen}
\end{figure}

We demonstrate this degeneracy in Figures~\ref{fig:crosscorr}, \ref{fig:tempvslong}, and \ref{fig:changeneigen}. Figure~\ref{fig:crosscorr} shows the cross-correlation coefficients at a wavelength of 2.41~$\mu$m for models using four (top panel) and five (bottom panel) eigencurves. When four eigencurves are used, the only terms showing high correlation coefficients $>0.8$ are the first two terms, which as described in \citet{Rauscher2018} are not expected to be orthogonal. However, with five eigencurves the other coefficients begin to show significant cross-correlation. The coefficients are designed to be orthogonal, so this correlation is a sign of using too many coefficients.

\citet{Rauscher2018} also found that using more eigencurves than can be well constrained by the data resulted in larger uncertainties on the derived temperature as a function of longitude. Figure~\ref{fig:tempvslong} shows that we find the same result for our test. This figure shows brightness temperature as a function of longitude for 1000 random samples from the MCMC fits using four vs. five eigencurves. With five eigencurves, the samples show a wider spread in derived temperatures. Additionally, when using five eigencurves, there are some samples where the temperature drops to unphysical negative values at some longitudes. This increase in the variance of the temperature with a larger number of eigencurves is due to the way in which each eigencurve coefficient is influenced by the information contained in the simulated observations vs. the prior on that coefficient. As described by \citet{Rauscher2018}, the first few eigencurves contain the most information from the data. In this case, the first four eigencurves have posteriors that are primarily driven by the data. However, the posterior of the fifth eigencurve is primarily driven by its prior and not the data. We use uninformative, uniform priors for the eigencurves, so when we add an eigencurve with a posterior driven by this uniform prior to our fit, it results in a much larger uncertainty in the fit.

This increased uncertainty from using too many eigencurves can also be seen in Figure~\ref{fig:changeneigen}, which compares the output eigenspectra when using four vs. five eigencurves. Even when the map is grouped into two regions in both cases, using four eigencurves results in both eigenspectra being better constrained and having smaller error bars than using five eigencurves.

When selecting the number of eigencurves to use to model a data set, we recommend using the method of \citet{Rauscher2018} and using the largest number of eigencurves for which none of them are significantly correlated with each other. We also found that adding additional eigencurves beyond this point resulted in an increase in the Bayesian Information Criterion (BIC), so this fitting criterion can be used to identify how many eigencurves to use.

One way to include more eigencurves in the fit would be to use more informative priors on the eigencurves. In this case, the priors could be selected based on expectations from a GCM or other model. While this could in principle permit a map that shows smaller-scale structures than our maps which use only the first four eigencurves, we choose to limit ourselves to considering fits with smaller numbers of eigencurves because we aim to determine how much information could be extracted from the data without incorporating any prior information from preexisting models. Additionally, while incorporating more restrictive priors could allow the use of more eigencurves without resulting in nonphysical solutions with negative temperatures, observing negative fluxes is informative because it shows that the solutions fall in a nonphysical region of parameter space.

\subsection{How many groups should be used?}
\label{sec:numgroups}

\begin{figure*}
\begin{center}
\begin{tabular}{cc}
\includegraphics[width=0.99\columnwidth]{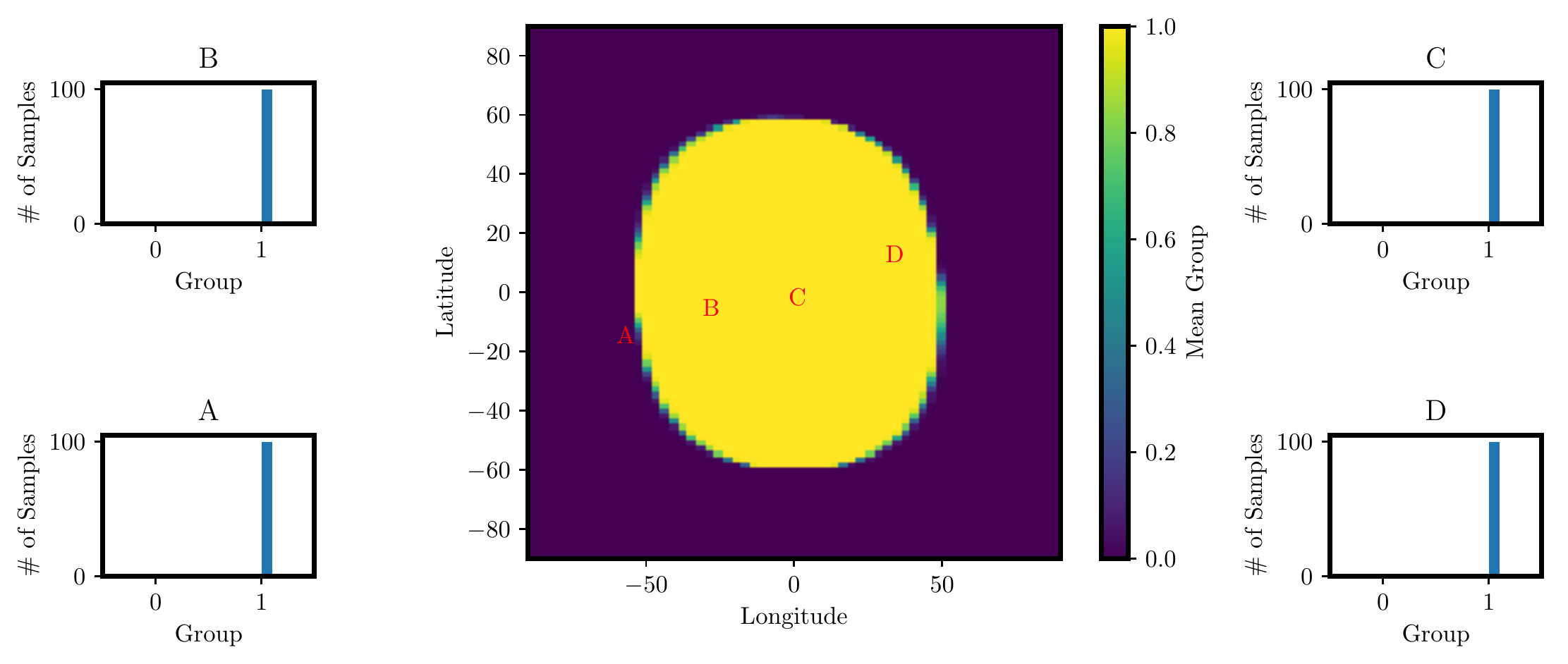} &
\includegraphics[width=0.99\columnwidth]{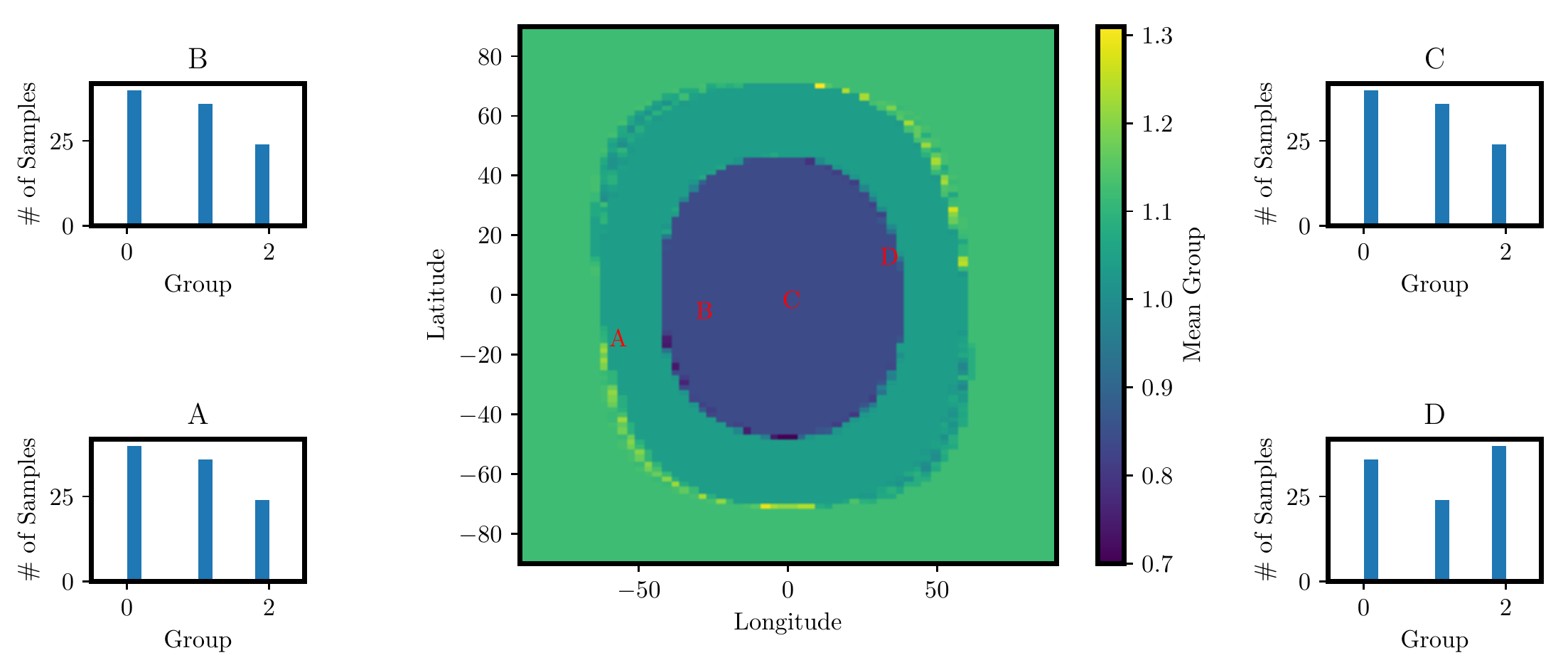} \\
\end{tabular}
\caption{Group assignments for the Continuum Hotspot model for two groups (left) and three groups (right). Maps show the mean group assignment across 100 realizations from the MCMC chain, and histograms show the full distribution of groupings across those 100 realizations for the four points indicated by red letters on each map. When two groups are used, almost all points in the map are consistently placed in the same group across all 100 realizations. However, with three groups there is much more variation in the grouping, which indicates that our simulated data are only precise enough to constrain two groups.}
\label{fig:2vs3groups}
\end{center}
\end{figure*}

We also used the Continuum Hotspot model to determine how many groups should be used when mapping. While the model contains 10 groups, the size of the error bars in the simulated observations determines how effectively these groups can be distinguished from each other. Figure~\ref{fig:2vs3groups} compares the planet maps when clustering the map into two or three groups. We performed the k-means clustering on 100 realizations of the map from the MCMC chain. The maps shown in Figure~\ref{fig:2vs3groups} display the mean group number at each point. The histograms show, for specific points on the maps, the grouping of that point over all of the MCMC samples which were run through the clustering algorithm.

We find that, for the Continuum Hotspot map, grouping the map into two groups results in clearly-defined groups, and at each step in the MCMC chain the group division occurs at almost the same position on the map. When using three groups, the mean map shows the shell structure contained in the original input map. However, the histograms show much more variation in the grouping of each individual point along the MCMC chain. With two groups almost all of the map pixels are consistently grouped into the same group, but with three groups the grouping varies significantly. This suggests that our data are only precise enough to constrain two distinct groups. We show the eigenspectra for the two-group case in Figure~\ref{fig:outputspec_continuum}.

When selecting the number of groups to use in the k-means clustering, we recommend using the largest number of groups such that most of the map pixels are still precisely constrained to be within a single group across the MCMC chain.

\subsection{Limits of Mapping Asymmetric Planets} \label{sec:asymmetric}

\begin{figure*}
\begin{center}
\includegraphics[width=0.99\textwidth]{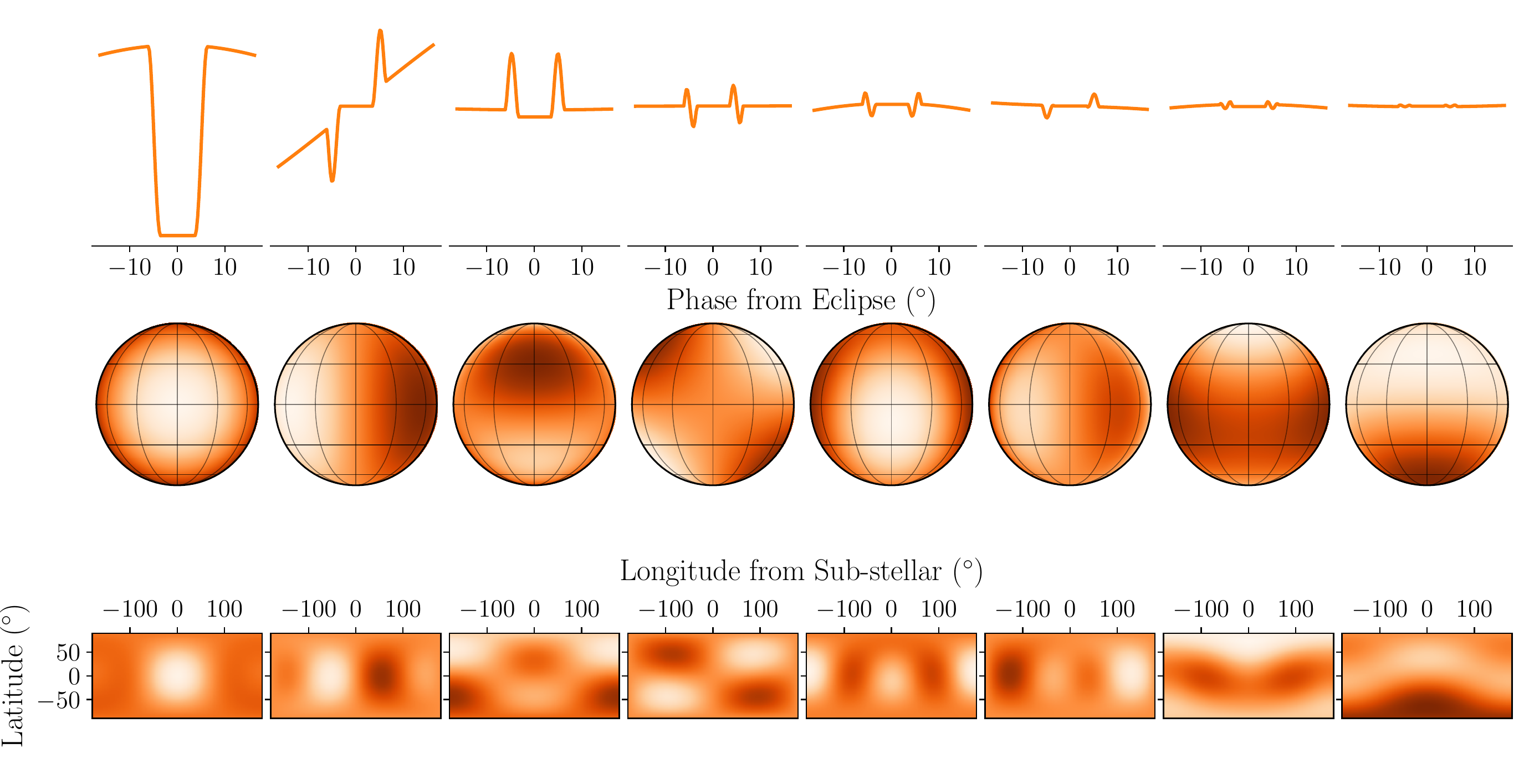}
\caption{Eigencurves and eigenmaps for the Asymmetric Hotspot Map. The top row shows the eigencurves, or the relative flux contribution as a function of phase from eclipse. The middle and bottom rows show two different projections of maps corresponding to each individual eigencurve. The final map was constructed from between 4-7 eigencurves depending on the wavelength, and the number of eigencurves at each wavelength was chosen following the procedure in Section~\ref{sec:numeigencurves}.}
\label{fig:eigencurves+maps}
\end{center}
\end{figure*}

The maps which have been discussed up to this point all show some form of a hotspot centered on the substellar point. However, close-in exoplanets may show hotspots offset from the substellar point \citep[e.g.][]{knutson2007map189,majeau2012eclipsemap189}, so we used the Asymmetric Hotspot map to examine how well a spherical harmonic-based model can represent structure on a map that is asymmetric about the substellar point. 
The eigenmapping method correctly identifies the location of the offset hotspot, as shown in Figure~\ref{fig:OffsetMaps}. However, one disadvantage of using the eigencurves method to model a planet map is they tend to produce structure that is somehow symmetric about the substellar point, because the first few eigencurves only give information on large-scale gradients that are all symmetric about the substellar point \citep{Rauscher2018}. This is demonstrated in Figure~\ref{fig:eigencurves+maps}, which shows maps corresponding to each individual eigencurve. The fourth map has a bright spot in the upper left quadrant of the dayside, similar to our Asymmetric Hotspot model, but also has a bright spot in the lower left quadrant. Combining this eigencurve with other eigencurves can mute the bright spot in the lower left quadrant slightly, but as shown in Figure~\ref{fig:OffsetMaps} the final map still shows a secondary hotspot in the lower left quadrant. Our clustering groups this secondary hotspot with the primary one, which causes mixing of the input spectra in the output eigenspectra (Figure~\ref{fig:outputspec_asymmetric}). Our method therefore seems to work best for maps which are symmetric about the substellar point or for determing large-scale flux gradients across the dayside, and small-scale structure within the maps should not be over-interpreted. However, our method is still useful for creating maps that 
only depend on a non-parametric model and are independent of any GCMs or other circulation models. More restrictive priors on the eigencurve coefficients could allow a fit with more eigencurves, which could in turn allow more accurate models of asymmetric flux distributions for the reasons discussed in Section~\ref{sec:numeigencurves}. However, for this paper we choose to examine what can be observed without incorporating prior information from GCMs.

\begin{figure*}
\centering
\includegraphics[width=\linewidth]{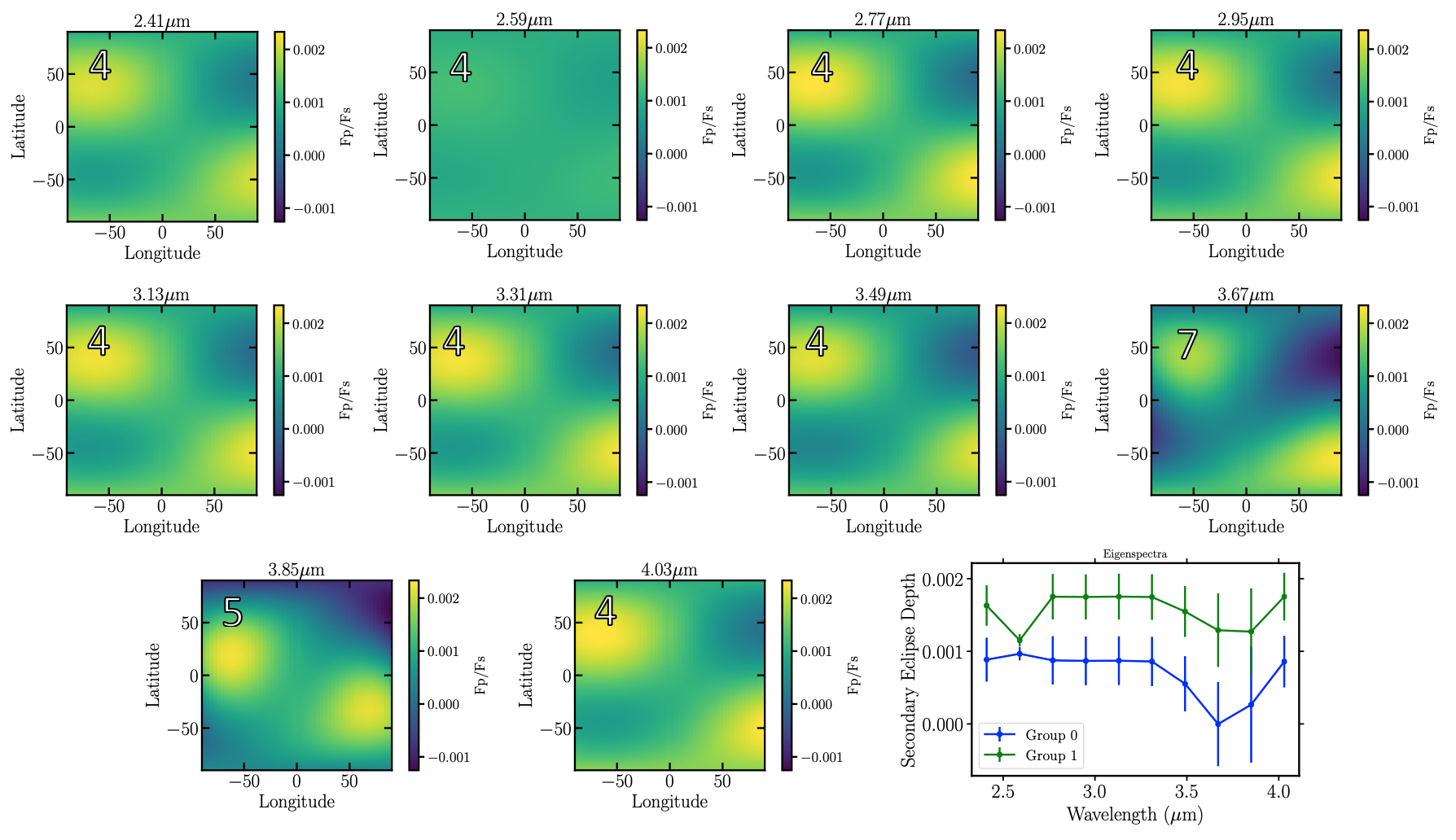}
\caption{Final output maps from the eigencurve fitting routine for each individual wavelength for the offset hotspot map. Note that the brightness scale in these maps shows the same mean intensity ratio as in Figures~\ref{fig:SpectralGroupMap}, \ref{fig:ContinuumMap}, and \ref{fig:OffsetMaps}. Numbers in the corner of each map show the number of eigencurves which was preferred at that wavelength. Bottom right plot shows the eigenspectra for the offset hotspot map. Generally the eigencurve fitting routine prefers more eigencurves at wavelengths where the difference between the eigenspectra is larger (e.g., 3.67~$\mu$m) and less eigencurves at wavelengths where the difference between the eigenspectra is smaller (e.g., 2.59~$\mu$m).}\label{fig:indeigenmaps}
\end{figure*}

Our results from the Asymmetric Hotspot model also reveal that our mapping method is more sensitive to planets where the flux gradient between the hottest and coldest points on the dayside is larger. Figure~\ref{fig:indeigenmaps} displays the median maps output at each individual wavelength, along with the final eigenspectra. Large numbers on each plot indicate the number of eigencurves that were favored at that wavelength based on the procedure described in Section~\ref{sec:numeigencurves}. We found that wavelengths where there is a larger contrast between the two input spectra allowed for a larger number of eigencurves to be fit, which in turn leads to a more detailed map.

\section{Conclusion}
\label{sec:conclude}

We have developed a framework to group the dayside emission spectra of exoplanets observed in secondary eclipse into unique spectral components (``eigenspectra'') emerging from different locations on the planetary disk. 
Our approach extends the method of 2D eclipse mapping into the wavelength dimension, opening the door to spatially-resolved studies of exoplanet atmospheric vertical structure and composition. To make such inference computationally tractable, our method identifies and groups spatial map components with intrinsically similar spectra, thereby reducing the dimensionality of any subsequent atmospheric retrieval.

Here we outline the main steps in our method. First, we use the \texttt{spiderman} package to fit a planet map to a light curve \citep{louden2017spiderman}. 
We base our fit on spherical harmonic maps and use the method of \citet{Rauscher2018} to construct orthogonal eigencurves from these spherical harmonic maps. Once we have constructed this map, we use K-means clustering to select regions of the map with similar spectra, and from each of these regions extract an ``eigenspectrum'', which is the mean spectrum of that region. These eigenspectra could then be analyzed with an atmospheric retrieval code to assess the chemistry and thermal structure of the planet.

We demonstrated how this method can be used to analyze multi-wavelength eclipse light curves of hot Jupiters using \textit{JWST}. To provide accurate and robust mapping results with our method, the following best practices should be used:
\begin{itemize}
    \item As was found in \citet{Rauscher2018}, the number of eigencurves used to construct a map at any given wavelength should be the largest number for which the coefficients to the eigencurves do not show any significant correlation. The number of eigencurves in a fit could be increased by using informative priors based on GCMs for the eigencurve coefficients.
    \item The number of unique spectra in the final map can be found by iterating the K-means algorithm until the recovered spectra are not overlapping and are separated into clearly defined regions of the map.
\end{itemize}

We additionally identify the following limitations of our method:
\begin{itemize}
    \item Structures that are strongly asymmetric about the substellar point are hard to fit well with eigencurves unless additional information from circulation models is incorporated into the fit.
    \item Using a finite number of eigencurves limits the spatial resolution of our map, so sharp gradients or discontinuities may be blurred. Our method should be able to recover that there \textit{are} large changes in conditions, but will not do a good job of resolving the \textit{spatial scale} of the change.
\end{itemize}

Our technique is readily able to identify that a planet has regions with distinct spectral features. However, it may not resolve the exact scale of features on planets with sharp spatial discontinuities in atmospheric structure or properties (e.g., aerosols, H$_{2}$O dissociation) or planets with multiple gradients simultaneously impacting the flux distribution unless they have particularly distinct regions. A general recommendation for using this method is that particular attention is given to which pieces of spectral-spatial information are or are not accessible in the observations. In particular, rather than presenting a derived map as the true ``image'' of the planet, the component parts that were used in the fit must also be shown, so that it is clear what was the potentially recoverable information.

However, hot Jupiters, which are the class of planet most amenable to eclipse mapping, are predicted to show large hemispheric gradients, which is the type of spatial pattern that the eigenspectra method could most easily map. Additionally, our method avoids assuming that the flux pattern across the planet follows expectations from any one physical model, making it a useful tool for first investigations of large-scale structure in a planet map regardless of the exact spatial-spectral patterns. More complex, physical models such as GCMs could be used to investigate the planet in more detail after the eigenspectra method was used to search for key large-scale patterns. The eigenspectra method could also be used to determine which features predicted by GCMs would be measureable via eclipse mapping.

With a large aperture and spectroscopic thermal infrared capability, \textit{JWST} promises precision data products capable of advancing the legacy of \textit{Spitzer}. 
In this paper, we have taken the first of many necessary steps towards a data-driven perspective on the 3D nature of exoplanet atmospheres.

\section*{Acknowledgements}

We thank the University of Michigan Institute for Research in Astrophysics for hosting the ``Multi-Dimensional Characterization of Distant Worlds" workshop in Ann Arbor during October 2018, where this project originated. MM acknowledges funding from a NASA FINESST grant. Funding for ES is provided by the NASA Goddard Spaceflight Center. JLY was supported by NASA's NExSS Virtual Planetary Laboratory funded by the NASA Astrobiology Program under grant 80NSSC18K0829. 

\textit{Software:} emcee \citep{foreman-mackey2013emcee}, HEALpix \citep{Gorski2005healpix}, matplotlib \citep{Hunter2007matplotlib}, numpy \citep{vanderWalt2011numpy}, scikit-learn \citep{pedregosa2011scikit-learn}, scipy \citep{Virtanen2019scipy}, spiderman \citep{louden2017spiderman}, starry \citep{luger2019starry}

\section*{Data Availability}

All data from this paper are publicly available on GitHub at https://github.com/multidworlds/eigenspectra.




\bibliographystyle{mnras}


\bsp	
\label{lastpage}
\end{document}